# Stochastic Approach to Turbulence: A Comprehensive Review

Ali Poursina [a], Ali Pourjamal [a], Ali Bozorg [a*]

a) Biotechnology Department, College of Science, University of Tehran, Iran

* Corresponding author:

Email: abozorg@ut.ac.ir

**Abstract**

Since Kolmogorov's theory, turbulence has been studied using various methods, many of which could be now be understood in a probabilistic framework. Herein, a comprehensive review of the advances made on stochastic theory of turbulence since Kolmogorov has been provided. It has been suggested that stochastic theory would be able to provide a natural foundation for a unified theory of turbulence as a wide range of problems related to turbulence could be treated using stochastic methods. At first, the mathematical theory of stochastic hydrodynamics has been studied and the results such as well-Posedness and ergodicity have been discussed for stochastic Burgers and Navier-Stokes equations. The theory of turbulence was then assessed where a special attention was paid to stochastic methods. Concepts including Burgers turbulence, scalar advection, incompressible turbulence, field theoretic methods, and stochastic variational principle have been studied to provide detailed discussion on the stochastic theory of turbulence. Moreover, an introduction to some stochastic numerical methods has also been provided. None of the covered methods have a counterpart in deterministic theories due to their derivation or inherent limitations of deterministic theories.

# 1. Introduction

## 1.1. An unsolved problem

Since Newton (1642-1727), history of science has witnessed a great deal of events and the most epic attempts have been made to achieve a more sophisticated insight into the nature. In such journey, with some successes and many failures, shifts of paradigms have taken place, ontological and phenomenological structures have been built or abandoned, formalisms have been established, and attempts for unification have been made. It is most certain that the chronological dynamics of science through the past few centuries has formed a complex unpredictable system. Complexity, in one of its most elegant forms, is the topic of this article.

Named by Leonardo da Vinci, turbulence have been a topic of interest that challenged many of the greatest minds in the course of history. Today, progress have been made, but nonetheless, turbulence remains far from being solved. The reason behind it, could be seen in the puzzling structure of the well-celebrated Navier-Stokes equation which determines the velocity field ($u$) in terms of time ($t$), pressure ($p$), density ($\rho$), and kinematic viscosity ($v$):

$$\begin{cases} \frac{\partial}{\partial t}\boldsymbol{u} + (\boldsymbol{u}.\nabla)\boldsymbol{u} = -\frac{\nabla \boldsymbol{p}}{\rho} + v\nabla^2 \boldsymbol{u} \\ \nabla.\boldsymbol{u} = 0 \end{cases} \quad (1)$$

Nowadays, we are almost aware how bizarre can a non-linear equation behave. Chaos, limit cycles, or bifurcations, are just a few of such strange behaviors. Navier-Stokes equation admits to its own specific non-linearity, which has turned it to one of the most formidable equations to tackle in mathematics. No global uniqueness and smoothness theorems have been proven for the three-dimensional Navier-Stokes equations yet and thereby, such incapability to explore the most basic properties of this equation implies that the advances in the mathematical theory of turbulence have not yet been sufficient for exploring turbulence in details.

As it was not possible to reconstruct turbulence from the first principles, a phenomenological description of turbulence was developed to better understand turbulence; Kolmogorov's theory, developed in 1941 [1], was partially successful in describing structure functions and energy spectrum of turbulent flows. This theory however, faces its own flaws; the two dimensional turbulence does not follow the Richardson cascade scheme, rather it follows a double cascade picture. Furthermore, in three dimensions, intermittency would also distort Kolmogorov's predictions on the structure functions.

There is no steadfast definition for complexity. However, it is a widespread and well justified belief that a turbulent dynamical system admits to complexity. The classical methods did not show much promise in describing complex systems. A major breakthrough in studying such systems was effectuated when a change of approach took place. To study complicated behavior of gas molecules, a system with a large number of degrees of freedom, probabilistic methods were employed by the founders of statistical mechanics. By doing so, although some information regarding the behavior of single molecules was lost since only averages were directly related to the observables but the obtained results were in a spectacular agreement with the general behavior of the system. However, a turbulent system is inherently dynamic and far from the equilibrium state, thus it could not be fitted in the structure of equilibrium statistical mechanics and new theories were needed to approach such problem.

As the central figure of turbulence theory, Kolmogorov also provided an axiomatic description of probability theory in 1933 [2] which facilitated further developments and resulted in the theory of stochastics that has been widely used in studying complex systems, ranging from neurons [3] to telecommunications [4] and the stochastic calculus provides a natural framework for studying continuous dynamical systems ranging from stock market to population dynamics in a probabilistic framework [5,6]. Addressing the turbulence with such methods has become a popular approach. Results from random dynamical systems, ergodic theory of stochastic differential equations (SDE), Langevin dynamics and Malliavin calculus were developed in order to solve scalar advection, two-dimensional turbulence, or other closely related problems to turbulence.

In this review, turbulence and mathematical hydrodynamics with proper stochastic tools would be discussed in detail. Figure 1 provides a structural scheme of this review. In Section 2 and 3, stochastic Burgers and Navier-Stokes equations would be first introduced and some of their essential features like existence, uniqueness, and ergodicity would be discussed. It should be noted that, although some results for weak solutions have been reported, no strong uniqueness theorem has been proved for three-dimensional case so far and ergodicity has been established only in a martingale sense. From another perspective, in Section 3, attempts to derive Navier-Stokes equation through stochastic variational principle would be presented and interesting results achieved by such approach would be summarized. Afterwards, In Section 4, turbulence would be discussed in details and Burgers turbulence, which share many of important properties of incompressible turbulence, would be studied. Field-theoretic methods would be introduced to estimate velocity gradient probability density function. It would also be shown that anomalies and spontaneous randomness would arise even in simplified models of scalar turbulence and thereby, inspired by quantum field theory, field-theoretic machinery would be applied to solve turbulent problems. Mathematical and physical description of multifractal models would also be presented and an alternative stochastic model for intermittency would be discussed briefly. Incompressible turbulence would be covered and dynamics, equilibrium, and non-equilibrium statistical mechanics of such systems would be studied in more detail. Moreover, existence of a unique random attractor for two dimensional Stochastic Navier-Stokes equation would also be discussed.

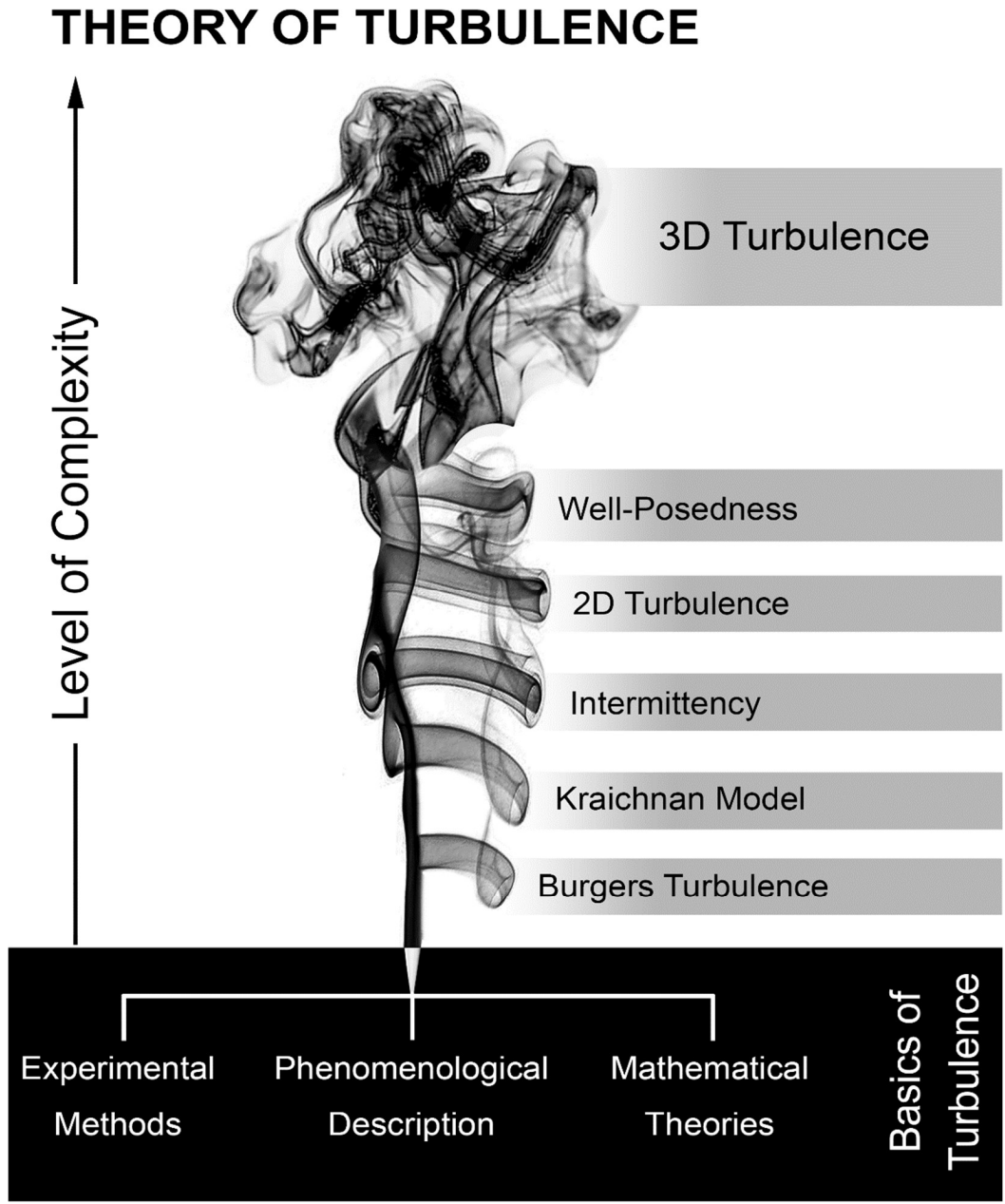

Figure 1: The theory of turbulence could be considered as tree where the roots are experiment, phenomenological theories such as Kolmogorov's theory, and first principles expressed based on mathematics. These models can be compared in terms of their complexity. Burgers model, could be considered as the simplest one-dimensional problem. However, as the number of dimensions increase, so does the complexity. Accordingly, one can think of well-Posedness and turbulence in three dimensions as the ultimate problems in this theory at the top of the turbulence tree.

Finally, to conclude the revision, numerical methods have been discussed in Section 5 to provide the readers with more comprehensive information and deeper understanding of methods which the stochastic theory can provide. A forward-backward system of SDEs for fluid flow would be summarized and Wiener Chaos Expansion (WCE) and its application in numerical methods would be specifically studied.

## 1.2. Scope

Turbulence has been a subject of interest in the last decades and its relevant topics have been widely investigated in the literature. However, a unified theory of turbulence is essential to understand, solve, and simulate the problems in this context. To date, despite all the efforts, the lack of such comprehensive theory has hindered further developments on turbulence and its wide application. In this review article, theoretical aspects of turbulence and the stochastic theory of hydrodynamics have been discussed in detail and thus, could be of benefit to the wide ranges of readers interested in understanding complexity and turbulence, including but not limited to physicists, mathematicians, engineers, or even biologists. Many of such information exploit advances in theory of probability and stochastic processes. Herein, a wide range of topics, from basics to advanced, have been covered where the most have been developed individually from fundamental expressions. Accordingly, by providing a map of theory of turbulence, provided information would be beneficial to those investigating the turbulence. Also, it should be noted that, mathematical theory of SDEs is based on functional analysis which might not be of interest to the non-mathematician readers. Therefore, in many cases, the strategies to prove a general result have been discussed and the readers have been referred to the literature for a complete proof.

It should be noted that, basic knowledge of hydrodynamics, Kolmogorov's theory, and the theory of probability would be required to follow the information summarized in this article. However, appendices, including a basic introduction to the theory of stochastic calculus, ergodic theory, and quantum field theory have also been presented to help readers better understand the provided information. Finally, although proved to be equivalent in some cases, it is necessary to mention that all the SDEs should be interpreted as Itô SDEs and not Stratonovich SDEs.

## 2. Burgers Equation

First introduced by Bateman in 1915 and developed by Burgers in 1948, Burgers equation not only provides scientists with a toy model of Navier-Stokes equation, but also describes phenomena in acoustic waves, cosmology, and traffic flow [7]. This equation, in the absence of external forcing, would be expressed as:

$$\frac{\partial u}{\partial t} + u \frac{\partial u}{\partial x} = \nu \frac{\partial^2 u}{\partial x^2} \qquad (2)$$

where $u$ is a given field and $\nu$ is a diffusion coefficient.

## 2.1 Deterministic Burgers Equation

Equation (2) is one of the rare non-linear partial differential equations(PDEs) that could be analytically solved for an explicit answer using the Hopf-Cole transformation [8,9] which reduces (2) to the heat equation in absence of a source:

$$\begin{cases} \psi(x,t) = -2\nu \ln \vartheta(x,t) \\ u(x,t) = \frac{\partial}{\partial x}\psi(x,t) \end{cases} \quad (3)$$

Therefore:

$$\frac{\partial}{\partial t}\vartheta(x,t) = \nu \frac{\partial^2}{\partial x^2}\vartheta(x,t) \quad (4)$$

Setting the initial condition $\psi(x,0) = \psi_0(x)$ then the solution can be expressed in terms of heat kernel:

$$\psi(x,t) = -2\nu \ln \int_{-\infty}^{\infty} \frac{dx_0}{\sqrt{4\nu t}} \exp\left(-\frac{(x-x_0)^2}{4\nu t} - \frac{1}{2\nu}\psi_0(x_0)\right) \quad (5)$$

$$u(x,t) = \frac{\int_{-\infty}^{\infty} dx_0 \frac{(x-x_0)}{t} \exp\left(-\frac{(x-x_0)^2}{4\nu t} - \frac{1}{2\nu}\psi_0(x_0)\right)}{\int_{-\infty}^{\infty} dx_0 \exp\left(-\frac{(x-x_0)^2}{4\nu t} - \frac{1}{2\nu}\psi_0(x_0)\right)} \quad (6)$$

Of particular interest would be the limit case of zero viscosity (i.e. $\nu \to 0^+$) with characteristic equation of:

$$\begin{cases} \frac{dx(t)}{dt} = u(x,t) \\ \frac{du(x,t)}{dt} = 0 \end{cases} \quad (7)$$

With the initial condition of $x(0) = x_0$, (6) could be solved for $x(t)$:

$$x(t) = x_0 + u(x_0, 0)t \quad (8)$$

It could be easily seen that, at $t_b = \min\limits_{x \in \mathbb{R}} \left[ \frac{-1}{\frac{du}{dx}} \right]$, two solutions with infinitely close boundary conditions would intersect each other. Therefore, to preserve uniqueness, solutions should also satisfy entropy condition (i.e. $u(x_+, t) \leq u(x_-, t)$) for all $(x, t)$.

To show that (5) satisfies (8) and shows discontinuity, (5) should be determined at the limit of zero viscosity. By using the steepest descent method [10], it could be concluded that:

$$\psi(x,t) = \max_{x_0 \in \mathbb{R}} [\psi_0(x_0) + \frac{(x-x_0)^2}{2t}] \tag{9}$$

$$u(x,t) = \frac{x - x_0(x,t)}{t} \tag{10}$$

It is obvious that (10) can be rearranged to (8). Besides, there exist two points corresponding to shocks with equal distance from $x$ that satisfy (9).

Finally, when $t \to \infty$ This scenario has been discussed by Hopf [9] as the following theorem:

*Theorem 1:*

*Let u(x,t) be a regular solution of Burgers Equation (2) for t>0*

*Define moment as $M = 2v \int_{-\infty}^{\infty} u(x,t) dx = 2vK$, where K is a constant*

*Let $\bar{x} = \frac{x}{\sqrt{2vt}}$, $\bar{u} = \sqrt{\frac{t}{2v}} u$, and $G(x) = e^{-K/2} \int_{-\infty}^{x} e^{-y^2/2} dy + e^{K/2} \int_{x}^{\infty} e^{-y^2/2} dy$*

*Then the following limit relation holds.*

$$\lim_{t \to \infty} \bar{u}(\bar{x}, t) = -\frac{G'(\bar{x})}{G(\bar{x})} \quad holds. \tag{11}$$

Meaning that the solutions to (2) converge uniformly to a stationary solution.

## 2.2 Adding Noise to Predictions

In this section, motivations behind using the stochastic differential equations (SDEs) would be discussed. Our understanding of the exact mechanisms behind natural phenomena is still flawed. Even the simplest real systems are subjected to many complicated interactions, and thereby, providing a complete model for all interactions is almost unfeasible. All the unaccounted interactions cause the state of the system to deviate from predictions given by a deterministic model. However, a possible solution could be the use of SDEs; SDEs can generate many realizations of the solution, which their mean value often corresponds to the prediction given by the deterministic model. However, a virtue of a well-devised stochastic model is to cover many possible scenarios, including extreme scenarios which are less probable to happen but in case of happening can have drastic consequences for the system. Stochastic models provide an opportunity to anticipate the worst-case scenarios, and moreover, help us to find a statistical description for the system based on realizations.

In general, in a complex system, where the non-linear dynamics rule, a simple perturbation can drastically change the future of the system. However, such perturbations do not exist in the realm of deterministic models. Turbulence is known to be unpredictable, and predictions of averaged models can be unreliable in some cases. For instance, it is not unusual to see velocity fluctuations about 25 % of the mean velocity in a wind tunnel and thereby, one could argue that there is more going on in the fluctuations than in the mean flow. Therefore, a stochastic model could reveal more about behavior of a complex system than its deterministic counter part.

Another topic of interest in hydrodynamics is the well-posedness of equations in fluid mechanics In some cases, solution to a PDE is not unique. The inviscid Burgers equation, before applying entropy condition, would be one of such PDEs of non-unique solution as we saw in previous section. However, it can be the case that the non-unique solutions would be unstable in the infinite dimensional space of the diffeomorphisms group which contains all the solutions to the PDE. In such cases, uniqueness by noise is a possible scenario[11]. The intuition behind this phenomenon is simple; as non-unique solutions are unstable in many cases, a random perturbation would restore the stability and uniqueness. In some other cases, singularities may arise in the solutions. In fluid dynamics such phenomenon could be seen in various solutions of Euler equation [12]. Random noises could also interact with these singularities in SDEs. However, prior to discuss regularization by noise in context of hydrodynamics, one should ask that whether regularization by noise would be possible or not.

### 2.3. Stochastic Burgers equation

### 2.3.1. Existence and uniqueness

Burgers equation with stochastic forcing could be expressed as:

$$\frac{\partial}{\partial t}u + u\frac{\partial}{\partial x}u = \nu \frac{\partial^2}{\partial x^2}u + f \qquad (12)$$

where $f(x,t)dt = \sum_k f_k(x)\, dW_k(t)$ is an additive noise and $\{W_k(.)\}$ are independent Wiener process. Assume homogeneous periodic boundary conditions on $R/Z = S^1$ and $f_k(\cdot) \in C^3(S^1), \left|\frac{\partial f_k}{\partial x}\right| \leq \frac{C}{k^2}$ denote probability space by $(\Omega, \mathscr{F}, \mathbb{P})$ and the filtration $\mathcal{F}_t$ generated by Wiener processes up to time t.

To determine the existence and uniqueness, (12) should be first written in a differential form:

$$du = \left(v\frac{\partial^2}{\partial x^2}u - \frac{1}{2}\frac{\partial}{\partial x}u^2\right) + \Sigma_k f_k(x)dW_k(t) \tag{13}$$

In a simplified case, (13) could be written as:

$$du = \left(v\frac{\partial^2}{\partial x^2}u - \frac{1}{2}\frac{\partial}{\partial x}u^2\right) + dW(t) \tag{14}$$

Define the self-adjoint operator A on $L^2(S^1)$:

$$Av = \frac{\partial^2}{\partial x^2}v \tag{15}$$

And consider the equation

$$dv = Av\, dt + dW(t) \tag{16}$$

Now obtaining the solutions to (17) would be straight forward and *u(t)* could be expressed in the form of:

$$v(t) = v_0 e^{tA} + \int_0^t e^{(t-s)A} dW(s) = v_0 e^{tA} + W_A(t) \tag{17}$$

where $v_0$ is the initial condition for (17).

Considering $v(t) = u(t) - W_A(t)$ with an initial condition of $v(0) = u_0$, (14) could be rewritten as:

$$dv = \left(Av + \frac{1}{2}\frac{\partial}{\partial x}(v + W_A)^2\right) dt$$

After such rearrangement, we would have:

$$v(t) = u_0 e^{tA} + \frac{1}{2}\int_0^t \frac{\partial}{\partial x}(v + W_A)^2 \, ds \qquad (18)$$

There are many versions of existence and uniqueness for stochastic Burgers equation [13–15]. This result has been proved for mild solutions using a fixed point argument by Da Prato and colleagues as following [16]:

*Theorem 2:*

*for $P \geq 2$ and $T^* > 0$ define:*

$$\Sigma_p(m, T^*) := \{v \in C([0, T^*]; L^p[0,1]): |v(t)|_{L^p[0,1]} \leq m \; \forall t \in [0, T^*]\}$$

*now for any any $m > |u_0|_{L^p[0,1]}$, there exists a stopping time $T^*$ such that (12) has a unique solution in $\Sigma_p(m, T^*)$*

Theorem 2 could be proved in 2 steps. At first, by defining

$$z(t) = u_0 e^{tA} + \frac{1}{2}\int_0^t \frac{\partial}{\partial x}(v + W_A)^2 \, ds = \mathcal{L}v$$

And using Sobolev embedding theorem, it can be shown that the following holds:

$$|z(t)|_{L^p[0,1]} \leq |u_0|_{L^p[0,1]} + \frac{1}{2}C(m + \mu_p)^2 (\frac{p-1}{2p}T^{*\frac{1}{2}-\frac{1}{2p}} + T^*) \text{ in which } \mu_p = \sup |W_A(t)|_{L^p[0,1]}$$

and $t \in [0, T^*]$ and $C$ is some positive constant. Therefore, there exists a $T^*$ such that $|z(t)|_{L^p[0,1]} < m$.

To prove uniqueness, assume $v_1, v_2 \in \Sigma_p(m, T)$ and $z_i = \mathcal{L}v_i$. Then set $z = z_1 - z_2$. Using Hölder's inequality and fixed point argument it could be shown that:

$$|z(t)|_{L^p[0,1]} \leq \frac{1}{2}C(m + \mu_p)(\frac{p-1}{2p}T^{*\frac{1}{2}-\frac{1}{2p}} + T^*)$$

So there exist a $T^*$ such that $\mathcal{L}$ is a strict contraction on $\Sigma_p(m, T^*)$. Moreover, it can be shown that if $v(t)$ in (18) is a bounded function from $[0, T]$ to $L^p[0,1]$, then $T^* = T$ which establishes existence and uniqueness of global solutions.

### 2.3.2. Ergodicity

In this section, ergodicity of Burgers equation would be discussed. Moreover, Supplementary Information (Appendix B) provides a basic discussion on of general theory of ergodicity. Proving ergodicity for an arbitrary system is not simple and in many cases, ergodicity would be assumed for the system without being actually proved. Turbulence was not an exception. Most the work on two dimensional turbulence in mid twentieth century was based on assumption of ergodicity without an actual proof. Ergodic measures are closely related to stationary and equilibrium states. Under some mild convergence assumptions, it has been proven by Sinai [17] that a unique invariant measure for viscid Burgers equation would exist on a torus.

*Theorem 3:*

*Let $u(x, 0) = \frac{\partial}{\partial x} \psi(x, 0)$, where $\psi$ is $C^2$ periodic function. Then the probability distribution $g_t$ of solution of (12), u(x, t), converges weakly to a steady state as t → ∞.*

To verify Theorem 3, an analogue to Hopf-Cole transformation using Itô calculus would be needed. For an explanation of Itô calculus please refer to the Supplementary Information (Appendix A).

Begin with heat equation in a random potential:

$$d\vartheta = \frac{\partial^2}{\partial x^2}\vartheta\, dt + (cdt - \frac{1}{2\nu}dW_t)\vartheta \tag{19}$$

take $\psi(x,t) = -2\nu \ln \vartheta(x,t)$, $c = \frac{1}{4\nu^2}$ and $u = \frac{\partial}{\partial x}\psi$. Therefore $u$ obeys:

$$\frac{\partial}{\partial t}u + u\frac{\partial}{\partial x}u = \nu \frac{\partial^2}{\partial x^2}u + \dot{W}_t \tag{20}$$

Solutions to (19) could be expressed through the Feynman-Kac formula as following:

$$\vartheta(x,t) = \mathbb{E}\left[\exp\left(\frac{1}{2\nu}\psi(x,0) - \frac{1}{2\nu}\int_0^t dW_t\right)\right] \tag{21}$$

To prove Theorem 3, (21) could be employed to define transfer matrices $T(x,t;x',t')$ for $\vartheta(x,t)$, where $x, x' \in S^1$ and $t' < t$. Using these matrices, it can be shown that:

$$u(x,t) = -2\nu \lim_{s \to \infty} \frac{\mathbb{E}\left[\frac{\partial}{\partial x}T(x,t;x_1,t-1)|x_s\right]}{\mathbb{E}[T(x,t;x_1,t-1)|x_s]}$$

which is measurable with respect to $\mathcal{F}_{t-s}^t$ (i.e., the filteration with intial time t-s at time t). This means that solutions to (20) display an independence from initial conditions and converge to a stationary solution at $t \to \infty$. For the complete proof visit [17]

Existence of invariant measure for inviscid case would be a more interesting question. Burgers system is continuously provided with energy. In the viscid case, a dissipation mechanism exists which maintains invariant distribution. On the other hand, the question arises whether or not entropy condition and shocks cause enough energy dissipation to preserve invariant measure in the inviscid limit. For this case, the following theorem was proven by Weinan *et al.* [18]:

*Theorem 4:*

*I) There exists a unique invariant measure $\mu_0$ for the inviscid case.*

*II) The invariant measure established for viscid case ($\mu_\nu$) converges weakly to $\mu_0$*

Write the inviscid Burgers equation as:

$$\frac{\partial}{\partial t} u + u \frac{\partial}{\partial x} u = f \tag{22}$$

where f has been introduced in (12).

The solutions to (22) are stochastic processes which admit jumps, and thus, Skorokhod space would be the most proper phase space for it. This space has been denoted by **D** and $\tilde{D}$ has been used to denote the borel σ-algebra on **D**. Due to the conservative nature of Burgers equation, without loss of generality, it could be assumed that $\int_{S^1} u(x,t) dx = 0$.

Define action of a Lipschitz curve $\xi: [t_1, t_2] \to \mathbb{R}$ as:

$$A_{t_1,t_2}(\xi) := \int_{t_1}^{t_2} \left[ \frac{1}{2} \left( \dot{\xi}(s) \right)^2 ds + \sum_k f_k(\xi(s)) dW_s \right]$$

Then solutions of (22) would satisfy:

$$u(x,t) = \frac{\partial}{\partial x} \inf_{\xi(t)=x} \left( A_{t_0,t}(\xi) + \int_0^{t_0} u(z, t_0) \, dz \right) \tag{23}$$

To establish a one-force one-solution principle, a one-sided minimizer could be defined as:

A one-sided minimizer is a $C^1$ curve $\xi \, (-\infty, 0] \to S^1$ if for any Lipschitz continuous curve $\eta: (-\infty, 0] \to S^1$ such that $\eta(0) = \xi(0)$ and η=ξ on $(-\infty, \tau]$ for some τ<0, then $A_{s,0}(\xi) \leq A_{s,0}(\eta)$ for all s ≤ τ.

if $\mathcal{M}_\omega = \{x \in S^1 : there\ exist\ more\ than\ one - sided\ minimizer\ such\ that\ \xi(0) = x\}$, then $\mathcal{M}_\omega$ is at most countable for almost every ω [19] and $u_\omega(x, 0) = \dot{\xi}(0)$. Based on the existence and uniqueness of solutions, solution operator as $S_\omega(t) u_\omega(x, 0) = u_\omega(x, t)$ could then be defined.

To define invariant measure, using Markov property of (22), the transition probability on **D** could be defined as:

$$P_t(u, A) = \int_\Omega \chi_A(u, \omega) \mathbb{P}(d\omega) \tag{24}$$

where $u \in \boldsymbol{D}, A \in \widetilde{D}$, and $if\ S_\omega(t)u \in A, \chi_A(u,\omega) = 1\ and\ otherwise, \chi_A(u,\omega) = 0$.

An invariant measure must satisfy the following relation on $(\boldsymbol{D}, \widetilde{D})$:

$$\mu_0(A) = \int_D P_t(u, A)\mu_0(du) \tag{25}$$

To prove the existence and uniqueness of (25), a shift operator could be defined as $\theta^\tau \omega(t) = \omega(t + \tau)$ and a mapping between $\omega$ and $u^\omega$ as $u^\omega = \Phi(\omega)$ to get $\Phi(\theta^\tau \omega) = S_\omega(\tau)\Phi(\omega)$. Now consider $\Phi_o: \Omega \to \boldsymbol{D}$ such that $\Phi_o(\omega)(x) = u^\omega(x, 0)$. It can shown that distribution of $\Phi_o$ is an invariant measure of the form (25) and existence is established with one-sided minimizers. For a proof of uniqueness, it could be shown that the existence of more than one invariant measure can lead to a contradiction with uniqueness of one-sided minimizers.

## 3. Navier Stokes Equation

In this section, topics on the existence, uniqueness, variational principle, and ergodicity of stochastic Navier-Stokes equation will be discussed.

Existence and uniqueness of the Navier-Stokes equation would be of interest in various contexts. Pioneered by Bensoussan and Temam [20], there exist a large amount of literature around this topic. Assuming $u(x, 0) \in H^1$, existence and uniqueness of strong local and global solutions for two [21] and three-dimensional [22] Navier-Stokes cases have been established. Existence and uniqueness of local solutions in Sobolev space $W_p^1(\mathbb{R}^d)$, where d>1 and p>d for a generalized Navier-Stokes equation [23], and local strong solutions in critical Besov spaces [24], have been established. While all the mentioned solutions hold in a strong probabilistic sense, solutions could also be interpreted in a weak sense(i.e. as martingales)[25,26].

### 3.1. Local strong solutions of stochastic Navier-Stokes equation

For an incompressible flow, equation of motion in stochastic form can be written as:

$$\begin{cases} \frac{\partial \boldsymbol{u}}{\partial t} + (\boldsymbol{u}.\nabla)\boldsymbol{u} = -\nabla p + \nu\nabla^2\boldsymbol{u} + \sum_k \boldsymbol{g}_k d\dot{W}_k \\ \nabla.\boldsymbol{u} = 0 \end{cases} \quad (26)$$

where $\boldsymbol{u}$ is defined on bounded domain $\mathcal{M}$ such that $\boldsymbol{u}|_{\partial\mathcal{M}} = 0$ and the existence of local strong solutions can be established.

*Definitions:*

*Define $H \coloneqq \{\boldsymbol{u} \in L^2(\mathcal{M}): \nabla.\boldsymbol{u} = 0, \boldsymbol{u}.\boldsymbol{n} = 0\}$ where H is a Hilbert space with inner product: $(u,v)_H = \int u.v \, d\mathcal{M}$ and n is unit normal vector of $\partial\mathcal{M}$.*

*For any separable Hilbert space H define $l^2(H)$ to be the set of all sequences $h = (h)_{k=1}$ for $h_k \in H$ such that $|h|_{l^2(H)} \coloneqq \sum_k |h_k|_H < \infty$.*

*Define $lip_u(X,Y)$ as the space of uniformly Lipschitz mappings from normed space X to normed space Y, $lip_u(X,Y)$ with conditions:*

*i)$|h(x) - h(y)|_Y \leq C|x - y|_X$*

*ii) $|h(x)|_Y \leq C(1 + |x|_X) \quad \forall h \in lip_u(X,Y))$*

*For some positive constant $C$.*

*Assume $g = \{g_k\}: \Omega \times [0, \infty) \times H \to l^2(H), g \in lip_u(H, l^2(H)) \cap C_0^\infty(\mathcal{M})^d$ where $d = 2,3$*

*Define $B(\mathbf{u}, \mathbf{w}) \coloneqq P_{div}(\mathbf{u}.\nabla)\mathbf{w}$, $A(\mathbf{u}) = P_{div}\nabla^2\mathbf{u}$, where $P_{div}$ is Leray projector and $A$ is known as Stokes operator with the domain $D(A) = H^2 \cap H_0^1$. It is well known that eigen-functions of this operator form an orthonormal basis $\{e_k\}$ with corresponding eigen values: $0 < \lambda_1 < \lambda_2 \leq \cdots \leq \lambda_n \leq \cdots$.*

*Define a local strong solution as $(u, \tau)$ where $\tau$ is strictly positive stopping time and $\mathbf{u}(. \wedge \tau) \in L^2(\Omega \times C[0,\infty); H)$ is a predictable process in H such that u satisfies the equation:*

$$\mathbf{u}(t \wedge \tau) + \int_0^{t\wedge\tau} \nu A\mathbf{u} + B(\mathbf{u}, \mathbf{u}) \, dt = \mathbf{u}_0 + \sum_k \int_0^{t\wedge\tau} \mathbf{g}_k dW_k \tag{27}$$

*Theorem 5:*

*There exist a unique local strong path-wise solution for a positive stopping time in 2 or 3 dimensions in sense of equation (27).*

Here, a sketch of proof has been provided and the readers are referred to the literature for a complete proof[26].

To prove the existence, Galerkin scheme would be used. By defining $H_n = span\{e_i\}_{i=1}^n$, (26) could be approximated as $u^n \in C([0,\infty); H_n)$ which would satisfy the following equations for all $v \in H_n$:

$$\begin{cases} d\langle \mathbf{u}^n, v\rangle + \langle \nu A\mathbf{u}^n + B(\mathbf{u}^n, \mathbf{u}^n), v\rangle = \mathbf{u}_0 + \sum_k \langle \mathbf{g}_k, v\rangle dW_k \\ \langle \mathbf{u}^n(0), v\rangle = \langle \mathbf{u}_0, v\rangle \end{cases} \tag{28}$$

This implies that the existence problem could be reduced to the existence of limit for each term in (28). Fortunately, for a positive stopping time $\tau$, each term in (28) would converge as below:

$$\mathbf{1}_{\Omega_{n'}} \mathbf{u}^{n'}(. \wedge \tau) \to \mathbf{u}$$

$$\mathbf{1}_{\Omega_{n'}} \int_0^{t \wedge \tau} \langle A\mathbf{u}^n, \mathbf{v} \rangle \, ds \to \int_0^{t \wedge \tau} \langle A\mathbf{u}, \mathbf{v} \rangle \, ds$$

$$\mathbf{1}_{\Omega_{n'}} \int_0^{t \wedge \tau} \langle B(\mathbf{u}^{n'}, \mathbf{u}^{n'}), \mathbf{v} \rangle \, ds \to \int_0^{t \wedge \tau} \langle B(\mathbf{u}, \mathbf{u}), \mathbf{v} \rangle \, ds \qquad (29)$$

$$\mathbf{1}_{\Omega_{n'}} \sum_k \int_0^{t \wedge \tau} \langle g_k(\mathbf{u}^{n'}), \mathbf{v} \rangle \, dW_K \to \sum_k \int_0^{t \wedge \tau} \langle g_k(\mathbf{u}), \mathbf{v} \rangle \, dW_K$$

where $\Omega_{n'} \uparrow \Omega$ and each converegence is a weak convergence in $L^2(\Omega \times [0, T])$. Using (29), it could be easily shown that $\mathbf{u}$ would satisfy (26).

In terms of uniqueness, it could be assumed that two solutions, namely $(\mathbf{u}', \tau)$ $and$ $(\mathbf{u}'', \tau)$, would exist with identical initial conditions:

$$\mathbb{P}(\mathbf{1}_\Omega \mathbf{u}_0' = \mathbf{1}_\Omega \mathbf{u}_0'') = 1 \qquad (30)$$

Now, to show uniqueness, one must prove that $\mathbb{P}(\mathbf{1}_\Omega \mathbf{u}'(t \wedge \tau) = \mathbf{1}_\Omega \mathbf{u}''(t \wedge \tau)) = 1$.

Assuming $\tilde{\mathbf{u}} = \mathbf{u}' - \mathbf{u}''$, inner product defined on $H$ could be used to write $|\mathbf{u}|^2 = (\mathbf{u}, \mathbf{u})_H$ and then use Ito formula to get:

$$d|\tilde{\mathbf{u}}|^2 = \frac{\partial |\tilde{\mathbf{u}}|^2}{\partial \tilde{\mathbf{u}}} d\tilde{\mathbf{u}} + \frac{\partial^2 |\tilde{\mathbf{u}}|^2}{\partial \tilde{\mathbf{u}}^2} (d\tilde{\mathbf{u}})^2 \qquad (31)$$

and

$$d\tilde{\mathbf{u}} = -\left(\nu A\tilde{\mathbf{u}} + B(\mathbf{u}', \mathbf{u}') - B(\mathbf{u}'', \mathbf{u}'')\right) dt + \sum_k \left(g_k(\mathbf{u}') - g_k(\mathbf{u}'')\right) dW_k \qquad (32)$$

Using definition of inner product and substituting (32) in (31), one can obtain:

$$d|\tilde{u}|^2 = [-2\nu((\nabla\tilde{u}, \nabla\tilde{u})_H) - 2\langle B(u', u') - B(u'', u''), \tilde{u}\rangle + 2\sum_k \langle g_k(u') - g_k(u''), \tilde{u}\rangle] dt +$$

$$\sum_k |g_k(u') - g_k(u'')|^2 \, dW_k \qquad (33)$$

For positive $R$, define stopping times as $\sigma_R := \inf_{t>0}\{(\nabla u', \nabla u')_H > R\} \wedge \tau$. After setting upper bound for $\langle B(u', u') - B(u'', u''), \tilde{u}\rangle$ [26] and using Lipschitz property of $g_k$, it could be shown by integrating (33) that:

$$\mathbb{E}\mathbf{1}_\Omega \left[|\tilde{u}(\sigma_R \wedge t)|^2 + \nu \int_0^{\sigma_R \wedge t}(\nabla\tilde{u}(s), \nabla\tilde{u}(s))_H \, ds\right] < C\, \mathbb{E}\mathbf{1}_\Omega \left[\int_0^{\sigma_R \wedge t}|\tilde{u}(s)|^2 \, ds\right] \qquad (34)$$

Using Poincare inequality, (43) results in:

$$\mathbb{E}\mathbf{1}_\Omega[|\tilde{u}(\sigma_R \wedge t)|^2] = 0$$

By simple calculations and assuming $R \to \infty$, it could be shown that $\mathbf{1}_\Omega(|\tilde{u}(t \wedge \tau)|^2) = 0$ which proves (30).

Existence of global solutions has been established only in two-dimensions. A global solution to (26) is the limit of its local solutions when $\tau \to \infty$. Considering $u_0 \in L^p(\Omega; H), p \geq 4$, the strategy is to define $\{\tau < \infty\} = \bigcup_T \{\tau \leq T\} = \bigcup_T \bigcap_n \{\rho_n \leq T\}$, where $\{\rho_n\}$ is an increasing sequence of stopping time. Since $\{\rho_n\}$ is increasing, it can be concluded that:

$$\mathbb{P}\left(\bigcap_n \{\rho_n \leq T\}\right) = \lim_{n\to\infty} \mathbb{P}(\rho_n \leq T)$$

Therefore, to prove the existence of global solutions, it would be enough to show that, for any fixed $T < \infty$, $\lim_{n\to\infty} \mathbb{P}(\rho_n \leq T) = 0$.

Defining $\gamma^M := \inf_{t \geq 0}\{\int_0^{t \wedge \tau} |u|^2 |\nabla \boldsymbol{u}|^2 \, ds > M\} \wedge 2T$, it can be shown that

$\lim_{n \to \infty} \mathbb{P}(\rho_n \leq T) \leq \mathbb{P}(\gamma^M \leq T)$ and $\mathbb{P}(\gamma^M \leq T) \leq \frac{1}{M} \mathbb{E}(\int_0^{T \wedge \tau} |u|^2 |\nabla u|^2 \, dt)$. Based on the fact that $\lim_{M \to \infty} \frac{1}{M} \mathbb{E}(\int_0^{T \wedge \tau} |u|^2 |\nabla u|^2 \, dt) = 0$, the existence of global solutions would be confirmed.

Moreover, uniqueness of global solutions in two-dimensional cases could also be affirmed. However, as the main idea behind such proof just resembles the local case, details of such proof would not be discussed here. Detailed proof with full discussion has already been provided by Mikulevicius and Rozovskii[27].

It should be noted that the proof of global existence could not be extended to three-dimensions. The reason behind it is that in (27), the upper bounds of operator $B$ are dimension-dependent and thus, have different expressions in two and three dimensions.

## 3.2. Ergodicity

Ergodic theory of stochastic Navier-Stokes equation has been a subject to many papers in the past two decades[28–32]. This theory is quite remarkable in various aspects. Physically, this theory facilitates statistical approaches to turbulence. In addition, an ergodic measure can act as a bridge between dynamics and phenomenological laws of fluids. Mathematically, the methods used in analysis of such theory consist of a wide range of topics in probability and infinite dimensional analysis, such as those employed in transition semi-groups, Malliavin calculus, and the theory of stochastic partial differential equations. A comprehensive and thorough review of Ergodic theory of hydrodynamics and the advances made in the past decades could already be found in the literature[33].

Here, some important results regarding the existence and uniqueness of invariant measures for (26), the equation of motion in stochastic form, will be briefly reviewed. However, it should be noted that the sophisticated mathematical methods behind these results would not be discussed in full details.

The uniqueness of global solutions for two-dimensional stochastic Navier-Stokes system was demonstrated. Taking the advantage of this property, transition semigroups could be used to establish invariant measures. a transition $(p_t)_{t>0}$ semigroup admits to the following properties:

i) $p_t\xi(x) = \mathbb{E}[\xi(u(t,x))]$ , $\xi \in \mathcal{B}(H), t \geq 0, x \in H$

in which $\mathcal{B}(H) = \{\psi: H \to \mathbb{R} \text{ where } \psi \text{ is a Borel function}\}$

ii) $p_{t+s} = p_t \circ p_s$

As could be inferred, the property $i$ is related to the existence of solution to (26) and property $ii$ depends on the uniqueness.

Considering that H is a Hilbert space, then $p_t\xi(x) = \int \xi(y)\, \mu_{t,x}(dy) = \mathbb{E}[\xi(u(t,x))] = \langle \xi, \mu_{t,x} \rangle = \langle p_t\xi, \mu_0 \rangle$ where $\mu_0$ is the intial measure for solutions to (26). This would result in $p_t^*\mu_0 = \mu_{t,x}$ where $p_t^*$ is the adjoint of $p_t$.

Now, a natural definition of invariant measure arises:

μ is an invariant measure if: $p_t^*\mu = \mu$

Probably the strongest result regarding two-dimensional case has been introduced by Hairer and Mattingly [34]:

*Definitions:*

*a transition semi-group is asymptotic strong Feller at $x \in H$ if following conditions are satisfied:*

*i) Assume there exist a sequence of pseudometiric $(d_n)_{n\in\mathbb{N}}$ such that $\lim_{n\to\infty} d_n(x,y)$*

$$= 1 \text{ if } x \neq y$$

*ii) There exist sequence $(t_n)_{n\in\mathbb{N}}$,*

$$t_n > 0 \text{ such that } \lim_{r\to 0} \limsup_{n\to\infty} \sup_{y\in B_r(x)} |P_{t_n}(x,.) - P_{t_n}(y,.)|_{d_n} = 0$$

*where $P_{t_n}(x,.)$ is defined in (24) and $|\mu|_{d_n} = \sup_{\lambda_{d_n}\psi=1} \int \psi(x)\mu(dx)$, $\lambda_{d_n}\psi = \sup_{\substack{x,y\in H \\ x\neq y}} \frac{\psi(x)-\psi(y)}{d_n(x,y)}$*

*Let $(\Omega_x, \mathcal{F}_x, \mathbb{P}_x)_{x\in D(A)}$ be a family of probability spaces and $(u(.,x))_{x\in D(A)}$ be a family of random prosseces on these spaces. The family $(\Omega_x, \mathcal{F}_x, \mathbb{P}_x, (u(.,x))_{x\in D(A)}$ is a Markov family if the following conditions hold:*

*i) $\mathbb{P}_x(u(t,x) \in D(A)) = 1 \quad \forall x \in D(A), t \geq 0$*

*ii) For each $F \in \mathcal{F}$, mapping $x \to \mathbb{P}_x(F)$ is measurable for any $x \in D(A)$*

*iii) for a borelian subsets of $\mathcal{A} \subset D(A), \mathbb{P}_x(u(t+s,x) \in \mathcal{A} \mid \mathcal{F}_s)$*

$$= \mathbb{P}_x(u(t+s,x) \in \mathcal{A} \mid u(s,x))$$

*And the Markov semi-group associated to this family is defined as:*

$$P_t\psi(x) = \mathbb{E}[\psi(u(t,x))] \quad x \in D(A), t \geq 0$$

*Theorem 6: if $(p_t)_{t>0}$ is asymptotic strong Feller at $x \in H$ then $x$ is in the support of at most one invariant measure.*

*Theorem 7: if $(p_t)_{t>0}$ is Feller and there exists random variable $u_0$ and $(t_n)_{n\in\mathbb{N}} \uparrow \infty$ and a probability measure $\mu$ such that:*

$$\frac{1}{t_n} \int_0^{t_n} \mu_{s,u_0} ds \to \mu$$

*where the convergence is a weak convergence, then $\mu$ is an invariant measure for $p_t$.*

By using Theorem 7, one can build an invariant measure and to prove the uniqueness, it would be enough to show $(p_t)_{t>0}$ is asymptotic strong Feller for every $x \in H$ and that 0 is in the support of every invariant measure.

This strategy has been used to establish ergodicity of two-dimensional stochastic Navier-Stokes equation with degenerate noise[34]. However, as a critical assumption, it was supposed that the white noise term in (26) had only two non-zero Fourier modes.

The non-degenerate case is simpler to work with and generalized results can be driven on a bounded domain. Since existence and uniqueness in two-dimensions are ensured, a Markov family associated with (26) could be built. Moreover, it can be shown that associated transition semi-group to the Markov family has a unique fixed point which would prove the general result when random force is non-degenerate.

As existence of a unique ergodic measure is related to stationary state, the rate at which the measure governing the system would converge to the stationary measure would be a topic of interest. To address this query, a distance measure for measures $\mu_i$ can be defined as:

$$\|\mu_1 - \mu_2\|_L^* := \sup_{f \in L_\mu^1} |\int f d\mu_1 - \int f d\mu_2|$$

Accordingly, it can be shown that for an arbitrary measure $\lambda$ and stationary measure $\mu$[35]:

$$\|p_t^*\lambda - \mu\|_L^* \leq C(|u|)e^{-\alpha t}$$

Due to uniqueness problem, treating the three-dimensional case proves to be a harder task and only partial results have been found so far. For instance, the strategy used by Da Prato and Debussche [36] for three-dimensional case is solving the Kolmogorov equation associated to stochastic Navier-Stokes equation and then demonstrate that the solutions to this equation correspond to the associated transition semi-group. Consider the stochastic Navier-Stokes equation in its abstract form:

$$\begin{cases} d\boldsymbol{u} + [vA\boldsymbol{u} + B(\boldsymbol{u},\boldsymbol{u})]dt = \sqrt{Q}\, dW_t \\ \boldsymbol{u}(0,.) = \boldsymbol{u}_0 \end{cases} \quad (35)$$

in which Q is the covariance operator. Therefore, the associated Kolmogorov equation can be represented as:

$$\begin{cases} \frac{\partial v}{\partial t} = \frac{1}{2} Tr(QD^2 v) - (vA\boldsymbol{u} + B(\boldsymbol{u},\boldsymbol{u}), Dv)_H \\ v(0,x) = \psi(x) \end{cases} \quad (36)$$

The Feynman-Kac semi-group $(S_t^m)$ can also be defined as:

$$S_t^m \psi(x) = \mathbb{E}\left[\exp\left(-K \int_0^t |A\boldsymbol{u}^m(s,x)|^2 ds\right) \psi(\boldsymbol{u}^m(t,x))\right] \quad (37)$$

where $\boldsymbol{u}^m$ satisfies Galerkin system of (35). Now the solution to (36) can be expressed as:

$$v = \lim_{m \to \infty} S_t^m \psi(x) + K \int_0^t |A|^2 v_m(s,x)\, ds \quad (38)$$

Subsequently, using the presented results, it would be possible to construct Markov solutions for (35) and to show close association between transition semi-group and solutions to Kolmogorov equation.

*Theorem 8: There exists a Markov family of martingale solutions to (35) and associated semi-group is continuous.*

It would be possible to build martingale solutions to three-dimensional Navier-Stokes solutions [37]. The main idea to prove ergodicity is to show, for each borelian function ψ on $D(A)$, $P_t \psi = v(t,x)$ which is a solution to (36). Such result facilitates the establishment of ergodicity for martingale solution of stochastic Navier-Stokes equation using strong Feller property of transition semi-group.

*Theorem 9 :[33] There exists a Markov process $(\mathbf{u}, \mu)$ on a probability space (Ω,ℱ, P) which is a martingale stationary solution of the stochastic Navier-Stokes equations (35) Where μ is the unique invariant measure of the transition semi-group$(p_t)_{t>0}$.*

Recall the close relation between martingale solutions and statistical conservation laws. It is a well-established observation that spontaneous symmetry breaking occurs in laminar to turbulent transition. However, in very high Reynolds numbers, symmetries would be restored in a statistical sense [38].

### 3.3. Stochastic Least Action Principle

In this section, a Lagrangian approach to hydrodynamics would be discussed. Pioneered by work of Arnold [39], this approach exploits group property of volume preserving diffeomorphisms of the configuration space to construct geodesics based on a kinetic energy Lagrangian. Arnold revealed that these geodesics could be used to build solutions to Euler equation. This approach led to some significant consequences including:

*i*) Volume preserving divergence free diffeomorphism group form a Lie group. In this manner, the results provided by Arnold could be considered as an extension Poincare's insight [40] into geometric mechanics [41]. Today, the general type of resulting equations are called Euler-Arnold equations.

*ii*) Studying the structure of geodesics facilitates stability analysis. Using generalized curvature for infinite dimensional spaces, one can deduce that geodesics with positive curvature are associated with converging fluid motions and laminar flows while negative curvature geodesics correspond to turbulence. A rather interesting result would also be a proof that curvature for diffeomorphism group associated to Euler equation could be strictly negative [41]. This implies that turbulence could exist in the limit of zero viscosity and in the lack of a dissipation mechanism.

In this section, a derivation to the Navier-Stokes equation using stochastic calculus will be studied. With some interesting mathematics, two approaches have been used for finding geodesics, one by using the Arnold's approach only in the context of stochastic calculus, and the other by using Lagrange multiplier method [42,43]. For a more detailed discussion, refer to the review conducted by Cruzeiro [44].

Consider the volume preserving space of diffeomorphisms on configurations space $\mathcal{S}$:

$Diff_V(\mathcal{S}) = \{g: \mathcal{S} \to \mathcal{S}; g \text{ is bijective } \& |\det D_t g| = 1 \ \& \ g \in L^2(\mathcal{S})\}$, where:

$$D_t g(t, f(x)_t) = \lim_{\varepsilon \to 0} \frac{1}{\varepsilon} E[g(t + \varepsilon, f(x)_{t+\varepsilon}) - g(t, f(x)_t)]$$

Define the following semi-martingale:

$$d\psi_t^u(x) = \sqrt{2\nu}\, dW_t + u(t, \psi_t^u(x))\, dt, \text{ where } \psi_t^u \in Diff_V(\mathcal{S}) \tag{39}$$

It should be noted that, for simplification, only one Weiner process has been used. However differential form of (39) could still be easily generalized linear combination of Weiner processes. Now, define the following action functional:

$$S[\psi^u] = \frac{1}{2} E\left[\int_0^T dt \int_S |D_t \psi_t^u(x)|^2 \, \mu(dx)\right] \text{ where } \mu \text{ is a measure defined on } S. \tag{40}$$

*Theorem 10: Define $\psi_t^u(x)$ that satisfies (39) where $u(t,.)$ divergence free vector feield on $[0,T] \times S$. Then $\psi_t^u(x)$ is a critical point of energy functional (40) if and only if $u(t,.)$ satisfies (1).*

Defining a variation of exponential form:

$$\begin{cases} exp_t(\eta q)(x) = x + \eta \int_0^t ds \, \dot{q}(s, exp_s(\eta q)(x)) \\ q(0) = q(T) = 0 \\ \nabla. q(t,.) = 0 \end{cases}$$

and minimizing the action functional (i.e., solving $\frac{d}{d\eta} S[\psi^u(.,\eta)]\Big|_{\eta=0} = 0$) leads to:

$$\int_0^T dt \int_S \left(\left(\frac{\partial}{\partial t} u + (u.\nabla)u - \nu \nabla^2 u\right).q\right)(t,x) \, \mu(dx) = 0 \tag{41}$$

Remember $\nabla. q(t,.) = 0$. So $\frac{\partial}{\partial t} u + (u.\nabla)u - \nu \nabla^2 u$ must be gradient of some function $p$ to satisfy (41) and theorem is proved: $(\frac{\partial}{\partial t} u + (u.\nabla)u - \nu \nabla^2 u = -\nabla p$.

This result is interesting in the sense that the deterministic Navier-Stokes equation could be obtained using a stochastic variational principle. In addition, as shown in Section 5, this result would also be used to introduce a system of SDEs that could be solved numerically in order to find the velocity field.

Furthermore, it is worth mentioning that this result could be extended to any Riemannian manifold [45]. By investigating the evolving nature of trajectories, Arnaudon and Cruzeiro also concluded that the trajectories would get apart exponentially and in the case of negative curvature and rotation of two particles would become more irregular over time [45]

## 4. Theory of Turbulence

In previous sections, the analytical theories of stochastic differential equations governing the fluid motion were discussed. Specifically, existence, uniqueness, and ergodicity of those equations were assessed and some results were derived. In this section, turbulence, a widely studied problem, especially within the last two centuries in mathematics, physics, and engineering, will be investigated (Fig. 1).

There is no doubt that Kolmogorov's insight provides more depth into turbulence and contributes to a better understanding of such unpredictable process. However, there has not been much progress in deducing the statistical description of Kolmogorov from first principles. Herein, the main objective is to study the theories of turbulence within the frame of recent advances in mathematical physics and experimental methods.

At first, Burgers turbulence, sometimes referred to as Burgulence, would be discussed and then a comprehensive study on scalar advection would be provided. Some features of Burgulence would later be seen to share basic properties of incompressible turbulence. As one of the few solvable models of turbulence, Kraichnan model has also been discussed as a toy model for scalar turbulence. Afterwards, intermittency has been studied both in physical and mathematical context. Incompressible turbulence, mostly in two dimensions, is the last topic of this section; methods of statistical mechanics and non-equilibrium thermodynamics have also been discussed and random attractors have been investigated in more detail.

In this chapter, as it serves better for the context of physical theories $\langle A \rangle$ would be used to denote average or expectation of function A.

## 4.1. Burgers Turbulence

A comprehensive study of Burgers equation and Burgulence has been conducted by Borichev and Kuksin [46]. Burgers equation proved to lack the required complexity for expressing features of incompressible turbulence. This can be easily verified by an explicit solution of Burgers equation derived by Cole and Hopf [8,9]. However, Burgulence shares some aspects of incompressible turbulence due to similar non-linearity in the governing equation.

To understand Burgers equation, structure functions are of particular interest:

$$S_n(\Delta x, t) := \langle |u(x + \Delta x, t) - u(x, t)|^n \rangle \tag{42}$$

For small $\Delta x$, sturcture functions could be approximated as [47]:

$$S_n(\Delta x, t) \sim \begin{cases} C_n |\Delta x|^n & n \in [0,1] \\ C'_n |\Delta x| & n \geq 1 \end{cases} \tag{43}$$

For Burgers system, recall $\frac{\partial}{\partial x}\psi(x,t) = u(x,t)$. It would be simple to verify that in limit of zero viscosity, the following relation holds for Burgers system:

$$\frac{\partial}{\partial t}\langle \psi(x,t) \rangle = E(t) \tag{44}$$

where $E(t) = \int \frac{1}{2} u^2(x,t) dx$.

To find a probability density function for $\psi(x,t)$, (9) could be then used to define a cumulative distribution function as follows:

$$F = \mathbb{P}\left\{\forall x_0 \ \psi_0(x_0) < \psi(x,t) + \frac{x_0^2}{2t}\right\} \tag{45}$$

For a crossing instance of $\psi_0(x_0) > \psi(x,t) + \frac{x_0}{2t}$ and considering the rarity of such instances, crossings could be assumed to follow a Poisson distribution [48]. Therefore:

$$\mathbb{P}\{no\ crossing\} = e^{-\langle N(t) \rangle} \tag{46}$$

where $\langle N(t) \rangle$ is mean number of crossing instances and can be written in an open form of:

$$\langle N(t) \rangle = \langle \int_{-\infty}^{\infty} \delta\left(\psi + \frac{x_0^2}{2t} - \psi_0\right)\left(\Theta\left(\frac{d\psi_0}{dx_0}\right)\right)\frac{d\psi_0}{dx_0} dx_0 \rangle$$

where $\Theta$ is the Heaviside function.

Now, by assuming a Gaussian distribution for $\psi(x, 0)$, this integral could be solved using Laplace's method [49]:

$$\lim_{t \to \infty} \langle N(t) \rangle = \sqrt{\frac{t}{\psi}} e^{-\psi^2} \tag{47}$$

Differentiating (46) with respect to $\psi$ leads to the probability density function $p(\psi)$. To find the maximum of this function, another differentiation should be carried out to get:

$\psi_P^2 \propto \ln(t)$ when $t \to \infty$ in the critical point P. thus:

$$\langle \psi(x,t) \rangle \propto \sqrt{\ln(t)} \tag{48}$$

$$E(t) \propto \frac{1}{t\sqrt{\ln(t)}} \tag{49}$$

$$l(t) \propto \left(\frac{t}{\ln(t)}\right)^{\frac{1}{4}} \tag{50}$$

Equation (50) could be considered as a scale of distance between shocks. Accordingly, as an interesting result, Considering the fact that $l(t)$ depends non-linearly on $t$, one can observe the effect of a long-term memory on Burgers system. As another interesting result, by rescaling the velocity with $\sqrt{E(t)}$ and distances with $l(t)$, it could be shown that the statistical properties of velocity approach a certain limit at $t \to \infty$, which would be in in line with ergodic theorems proved for Burgers equation.

For further discussion, a propagating shock-wave solutions of (2) would be of interest. These solutions could be derived when writing solutions of (2) as a traveling wave of $u(x,t) = v(x - st)$. Now it could be easily shown that $u = v \tanh\left(v \frac{x}{2\nu}\right)$ is a solution to (2) in which $v = \frac{u\left(-\frac{\Delta x}{2}\right) - u\left(\frac{\Delta x}{2}\right)}{2}$.

By defining $E_n(t) = \langle \int u^{2n}(t,x) \, dx \rangle$, a relation similar to that of Karman-Howarth relation in Navier-Stokes could be derived [50]:

$$\frac{\partial S_{2n}}{\partial t} = -\frac{2n-1}{2n+1} \frac{\partial S_{2n+1}}{\partial x} + \nu \frac{\partial^2 S_{2n}}{\partial x^2} - 4 \frac{\partial E_n(t)}{\partial t} \tag{51}$$

Since Burgers system is an ergodic system, by considering a steady state limit, the inviscid limit can be used to deduce from (51) that:

$$S_{2n+1} = -4 \frac{2n+1}{2n-1} \frac{\partial E_n(t)}{\partial t} \tag{52}$$

Subsequently, existence of non-zero odd moments could be inferred from (52) which would imply the breaking of time-reversal symmetry such that:

$$\mathbb{P}(\delta u, x) \neq \mathbb{P}(-\delta u, x)$$

In addition, this result would imply the non-Gaussian distribution of probability density function for velocity increments, which is a well-known result for incompressible turbulence.

Recall that Navier-Stokes and Burgers equation are invariant under the following transformation:

$$x \to \lambda x \quad u \to \lambda^h u \quad t \to \lambda^{1-h} t \quad \nu \to \lambda^{1+h} \nu \tag{53}$$

So, the energy dissipation rate can be scaled to:

$$\frac{\partial E(t)}{\partial t} \to \lambda^{3h-1} \frac{\partial E(t)}{\partial t} \sim \nu^{\frac{3h-1}{1+h}}$$

However, energy dissipation rate is independent of viscosity as it remains finite in limit of zero viscosity, phenomenon known as dissipative anomaly. In general, even in the absence of any interruption, anomaly arises when a symmetry is not restored. Accordingly, the only probable value for h would be $\frac{1}{3}$ in very large Reynold numbers, which would be one of the results of Kolmogorov's theory of turbulence.

Now, to investigate scale-invariance $(i.e., \mathbb{P}(\delta u, x) = \frac{1}{\delta u} P(\frac{\delta u}{x^h}))$, since $S_3 = -12x \frac{\partial E_2(t)}{\partial t}$, the scale-invariance of (53) suggests that $S_n \propto \left(x \frac{\partial E_2(t)}{\partial t}\right)^{\frac{n}{3}}$. However, this is not the case for propagating shock-wave solutions therefore, the scale-invariance symmetry breaks and as the scale decreases, the lack of such symmetry would lead to a faster decay of low–order moments than those of the high order ones and therefore, intermittency arises. This implies that, in general, $S_n \propto \left(x \frac{\partial E_2(t)}{\partial t}\right)^{\zeta(n)}$ and when $\zeta(n)$ is not linear with respect to $n$ which gives rise to anomalous scaling. Also, in terms of breakdown of Galilean invariance, it should be noted that neither $E_n$ nor $S_{2n+1}$ are Galilean invariant.

When considering anomalies, quantum field theory (QFT) comes to mind as anomalies are often encountered in that context. Inspired by methods of QFT, problems of statistical physics adopted field theoretic methods in face of such anomalies. Now, stochastic Burgers equation of form (12) in which f is a Gaussian random force could be assessed. The results derived here would be mostly based on the work of Polyakov [51] that is known as one of the first in which field theoretic methods were employed to study Burgers equation. The approach used by Polyakov was based on operator product expansion Assume:

$$\frac{\partial u}{\partial t} + u\frac{\partial u}{\partial x} = \nu\frac{\partial^2 u}{\partial x^2} + f(x,t)$$

Where $f(x,t)$ is white in time:

$$\langle f(x,t)f(y,s)\rangle = \delta(t-s)B(x-y) \tag{54}$$

And $B$ is the spatial correlation of random force. Now a n-point generating functional (i.e., characteristic function) can be defined as:

$$Z_n(\lambda,x,t) = \langle \exp(\sum_{i=1}^{n} \lambda_i u(x_i,t))\rangle \tag{55}$$

By doing so, it would be convenient to verify:

$$\dot{Z}_n + \sum_{i=1}^{n} \lambda_i \frac{\partial}{\partial x_i} Z_n = \mathcal{D}^\nu(\lambda,x,t;n) + \frac{1}{2}\sum_{i,j=1}^{n} B(x_i - x_j)\lambda_i \lambda_j Z_n \tag{56}$$

where in the R.H.S, the last term could be justified with Ito formula and Gaussian nature of random force and first term can be written as following:

$$\mathcal{D}^\nu(\lambda,x,t;n) = \nu \langle \sum_{i=1}^{n} \lambda_i \frac{\partial^2 u_i}{\partial x_i^2} \exp(\sum_{j=1}^{n} \lambda_j u(x_j,t))\rangle \tag{57}$$

As solutions of Burgers equation contain singular derivatives in limit of zero viscosity, the limit of (57) remains non-zero. Such anomaly inspires using field theoretic methods to calculate probability density functions of velocity increments. Here Polyakov introduces the conjecture [51] the operator product expansion is related linearly to $Z_n$ and preserves statistical symmetries of burgers equation. Therefore:

$$a_\lambda = \lim_{v \to o} v \frac{\partial^2 \lambda u(x)}{\partial x^2} \exp(\lambda u(x)) = \left(\frac{a}{2} + \frac{b-1}{\lambda}\frac{\partial}{\partial x} + c\lambda \frac{\partial}{\partial \lambda}\right) \exp(\lambda u(x)) \qquad (58)$$

For a two-point correlation function with these assumptions, the following equation can be derived from (58):

$$\frac{\partial^2 Z_2}{\partial \mu \partial y} - 2\left(B(0)\Lambda^2 + \frac{\partial^2 B(x)}{\partial x^2}\bigg|_{x=0}(\mu y)^2\right) Z_2 = aZ_2 + \frac{2b}{\mu}\frac{\partial Z_2}{\partial y} + c\Lambda \frac{\partial Z_2}{\partial \Lambda} + c\mu \frac{\partial Z_2}{\partial \mu} \qquad (59)$$

where a, b, and c are parameters determined only indirectly and considering the change of variables $\lambda_{1,2} = \Lambda \pm \mu$ and $x_{1,2} = x \pm \frac{y}{2}$.

Also, by considering $Z_2(\Lambda, \mu, y) = \Phi(\mu y)\Psi(\Lambda)$, (59) can be written as a closed system of t equations:

$$\begin{cases} -2B(0)\Lambda^2 \Psi(\Lambda) = c\Lambda \frac{\partial \Psi(\Lambda)}{\partial \Lambda} \\ \frac{\partial^2 \Phi(\mu y)}{\partial \mu \partial y} - 2\frac{\partial^2 B(x)}{\partial x^2}\bigg|_{x=0}(\mu y)^2 \Phi(\mu y) = a\Phi(\mu y) + \frac{2b}{\mu}\frac{\partial \Phi(\mu y)}{\partial y} + c\mu \frac{\partial \Phi(\mu y)}{\partial \mu} \end{cases} \qquad (60)$$

From (60), probability density functions for velocity increments can be deduced using two-point correlation function:

$$\lim_{\zeta \to \infty} p(\zeta) = \zeta^k e^{-C\zeta^3} \qquad (61)$$

$$\lim_{\zeta \to -\infty} p(\zeta) = |\zeta|^{-\tilde{k}} \qquad (62)$$

where C depends on the random force and $k = 2b - 1$, $\tilde{k} = 2b + 1$ and $\zeta = \frac{\partial u}{\partial x}$.

Although there have been some debates on the value of b, so far, the most favored value has been $b = \frac{5}{4}$. For further discussion, one is referred to the study conducted by Bec and Khanin [7] and the references within.

### 4.2. Scalar Advection

A reaction diffusion process can be represented by following equation:

$$\frac{\partial \Phi}{\partial t} + (u.\nabla)\Phi = D\nabla^2 \Phi + S(x,t,\Phi) \qquad (63)$$

where $S$ is the term related to production or consumption of $\Phi$. Scalar advection is an important problem for modeling transport phenomena and also two-dimensional turbulence as it is simple to verify vorticity in two dimensions obeys an equation similar to that of (63), but in the absence of the source term $S$.

In a realistic model, (63) must be coupled with (1) and value of $u$ could be determined with solving Navier-Stokes equation. However, if not impossible, an analytical solution for this system of equations is unattainable at the moment. In 1968, Kraichnan [52] developed a simplified model of scalar advection where the velocity was white in time. Such simplified approach could not only be used as a toy model to understand some of the properties of turbulent scalar advection, but could also be considered as a test ground for developing new tools in turbulence theory such as field theoretic models.

### 4.2.1. Kraichnan Model

To study scalar advection, a switch to Lagrangian approach (similar to section 3.3) would be beneficial. Denote the Lagrangian trajectory in time $t$ by $\boldsymbol{R}(t; t_0, \boldsymbol{x_0})$ with initial starting point $\boldsymbol{x_0}$ at time $t_0$ (equal to 0 unless stated otherwise). In Kraichnan model, such trajectories obey:

$$d\boldsymbol{R} = dW_t = \boldsymbol{u}\, dt \tag{64}$$

where:

$$\langle u_i(t,\boldsymbol{x}) u_j(t',\boldsymbol{x'}) \rangle = \delta(t-t') D_{ij}(\boldsymbol{x},\boldsymbol{x'}) \tag{65}$$

Furthermore, homogeneity, isotropy, and scaling property is also assumed as:

$$D_{ij}(\boldsymbol{x},\boldsymbol{x'}) = D_{ij}(|\boldsymbol{x}-\boldsymbol{x'}|) \tag{66}$$

$$D_{ij}(\boldsymbol{x}) = \delta_{ij} D'(|\boldsymbol{x}|) + \frac{x_i x_j}{|\boldsymbol{x}|^2} D''(|\boldsymbol{x}|) \tag{67}$$

$$D^{ij}(\boldsymbol{x}) = \begin{cases} \mathcal{O}(|\boldsymbol{x}|^2) & |\boldsymbol{x}| \ll \eta \\ \mathcal{O}(|\boldsymbol{x}|^\xi) & \eta \ll |\boldsymbol{x}| \ll L \end{cases} \tag{68}$$

where $\xi = \frac{4}{3}$ in Kolmogorov theory.

Now, the forward Kolmogorov equation of (64) could be derived:

$$\frac{d}{dt}\langle f(\boldsymbol{R}) \rangle = \frac{1}{2} D_{ij}(0) \langle \nabla_i \nabla_j f(\boldsymbol{R}) \rangle \tag{69}$$

From isotropy, $D_{ij}(0) = D_0 \delta_{ij}$ and then, the time evolution of probability density function could be found from (69):

$$\frac{d}{dt} P(t,\boldsymbol{x};\boldsymbol{x_0}) = \frac{1}{2} D_0 \nabla^2 P(t,\boldsymbol{x};\boldsymbol{x_0}) \tag{70}$$

Solution to (70) could be represented as heat kernel:

$$P(t, \boldsymbol{x}; \boldsymbol{x_0}) = \frac{1}{(2\pi D_0 t)^{\frac{d}{2}}} \exp\left(\frac{(x-x_0)^2}{2D_0 t}\right) \tag{71}$$

Hence:

$$\langle \Phi(t, \boldsymbol{x}) \rangle = \int \Phi(0, \boldsymbol{x_0}) \frac{1}{(2\pi D_0 t)^{\frac{d}{2}}} \exp\left(\frac{(x-x_0)^2}{2D_0 t}\right) d\boldsymbol{x_0} \tag{72}$$

Where $d$ is dimension of space.

To further analyze Kraichnan model, some tools from ergodic theory is required. A proper measure for stability could be defined as:

$$W_i^j := \frac{\partial}{\partial x_0^j} R_i(t; t_0, \boldsymbol{x_0}) \tag{73}$$

where 0 is the initial time leading to $\boldsymbol{R}(t_0; t_0, \boldsymbol{x_0}) = \boldsymbol{x_0}$. By using the chain rule, it could be simply verified that $W(t; t_0, \boldsymbol{x_0}) = W^{-1}(t_0; t, \boldsymbol{x})$.

Decomposing W results in:

$$W_\mu = \mathfrak{O}' diag(e^{\mu_1}, \dots, e^{\mu_d}) \mathfrak{O} \tag{74}$$

where $\mu_1 \geq \cdots \geq \mu_d \geq -\infty$.

Now, consider the limit $\lambda_i = \lim_{t \to \infty} \frac{\mu_i(t; x_0)}{t}$ in which $\lambda_i$ is called a Lyapunov exponent. The existence of finite Lyapunov exponent has been proved in general context of ergodic theory by Osedelec in 1965 [53] under mild assumptions which is known as Multiplicative Ergodic Theorem (MET).

At this stage, it would be possible to find an equation for dynamics of $W$ in Kraichnan model. Considering $\delta R$ as the distance between Lagrangian trajectories, $\delta R$ could be written in following Itô differential form:

$$\begin{cases} d\delta R = (\delta R.\nabla)u\, dt := s(t)dt\, \delta R \\ s_i^j(t) = \nabla_j u_i(t, R(t)) \end{cases} \quad (75)$$

Differentiating (75) with respect to $x_0$ would provide us with an SDE which governs dynamics of matrix $W$:

$$dW = s(t)dt\, W \quad (76)$$

where, $s(t)$ is a matrix valued noise with the correlation function of:

$$\langle s_k^i(t)s_l^j(t')\rangle = \delta(t-t')C_{kl}^{ij} \quad (77)$$

With a straightforward calculation, the following equation could be obtained:

$$\frac{d}{dt}\langle f(W)\rangle = \frac{1}{2}C_{kl}^{ij}\langle W_m^k W_n^l \frac{\partial^2}{\partial W_m^i \partial W_n^j}f(W)\rangle \quad (78)$$

Considering isotropy assumption, certain limitations could be applied to $C_{kl}^{ij}$ such that:

$$C_{kl}^{ij} = b(\delta_k^i \delta_l^j + \delta_l^i \delta_k^j) + c\delta^{ij}\delta_{kl} \quad (79)$$

To further develop the calculation, (79) could be inserted in (78) to get:

$$\frac{1}{2}C_{kl}^{ij}W_m^k W_n^l \frac{\partial^2}{\partial W_m^i \partial W_n^j} = \frac{1}{2}\left(bW_m^i \frac{\partial}{\partial W_m^i}W_n^j \frac{\partial}{\partial W_n^j} - bW_m^i \frac{\partial}{\partial W_m^i} + bW_m^j \frac{\partial}{\partial W_m^i}W_n^i \frac{\partial}{\partial W_n^j} - dbW_m^j \frac{\partial}{\partial W_m^j}\right.$$

$$\left. + cW_m^k \frac{\partial}{\partial W_m^i}W_n^k \frac{\partial}{\partial W_n^i} - cW_m^k \frac{\partial}{\partial W_m^k}\right)$$

which would result in:

$$\frac{d}{dt}\langle f(W)\rangle = \langle \left(\frac{c}{2}\sum_{i,j}(\mathcal{T}_j^i)^2 + \frac{b}{2}\sum_{i,j}\mathcal{T}_j^i\mathcal{T}_i^j + \frac{b}{2}\mathcal{L}^2 - \frac{(d+1)b-c}{2}\mathcal{L}\right)f(W)\rangle \tag{80}$$

where $(\mathcal{T}_i^j f)(W) = \frac{d}{d\varepsilon}f(e^{-\varepsilon E_i^j}W)$ and where $E_i^j$ has only one non-zero element on the intersection of j$^{th}$ column and i$^{th}$ row and $(\mathcal{L}f)(W) = \frac{d}{d\varepsilon}f(e^\varepsilon W)$.

Considering (74), (80) could then be expressed as:

$$(\mathcal{T}_i^j f)(W_\mu) = -\frac{\partial}{\partial \mu_i}f(\boldsymbol{\mu})$$

$$((\mathcal{T}_j^i)^2 f)(W_\mu) = \frac{\partial^2}{\partial \mu_i^2}f(\boldsymbol{\mu})$$

$$(\mathcal{L}f)(W_\mu) = \sum_i \frac{\partial}{\partial \mu_i}f(\boldsymbol{\mu})$$

$$(\mathcal{L}^2 f)(W_\mu) = \sum_i \left(\frac{\partial}{\partial \mu_i}\right)^2 f(\boldsymbol{\mu})$$

$$(\mathcal{T}_j^i \mathcal{T}_i^j f)(W_\mu) = \coth(\mu_i - \mu_j)\frac{\partial}{\partial \mu_i}f(\boldsymbol{\mu})$$

As a result, it could be concluded that:

$$\frac{d}{dt}\langle f(\boldsymbol{\mu})\rangle = \langle \mathcal{H}f(\boldsymbol{\mu})\rangle = \langle \frac{b+c}{2}\left(\sum_i \frac{\partial^2}{\partial \mu_i^2} + \sum_{i\neq j}\coth(\mu_i - \mu_j)\frac{\partial}{\partial \mu_i}\right)f(\boldsymbol{\mu}) + \frac{b}{2}\sum_i\left(\frac{\partial}{\partial \mu_i}\right)^2 f(\boldsymbol{\mu}) -$$

$$\frac{(d+1)b+c}{2}\sum_i \frac{\partial}{\partial \mu_i}f(\boldsymbol{\mu})\rangle \tag{81}$$

Now, using related Kolmogorov's equation:

$$P(\boldsymbol{\mu},t) = e^{t\mathcal{H}}(\boldsymbol{\mu}) \tag{82}$$

Using the approximation scheme introduced by Balkovsky and Fouxon [54] for (82), Lyapunov exponents could be calculated:

$$\lambda_i = \frac{cd}{2} - bi \tag{83}$$

Accordingly, (83) implies that chaotic phase would take place if $cd > 2b$.

Now, by considering Kraichnan model in inertial range $\Delta := |R_1(t) - R_2(t)|$ could be defined as the distance between two Lagrangian trajectories. Similar to equation (75) and considering the scaling law, it could be shown that [55]:

$$\begin{cases} \frac{d}{dt}\langle f(\Delta)\rangle = \langle \mathcal{M}f(\Delta)\rangle = \frac{2b+c}{2}\Delta^{\xi-a}\frac{d}{d\Delta}\left(\Delta^a \frac{d}{d\Delta} f(\Delta)\right) \\ a = \frac{(d-1)c}{2b+c} \end{cases} \tag{84}$$

and

$$P(\Delta, t; \Delta_0) = e^{t\mathcal{M}}(\Delta; \Delta_0) \tag{85}$$

In an interesting case in which $\mathcal{M}$ exhibits reflecting boundary condition of $\left(\Delta^a \frac{d}{d\Delta} f\right)(0) = 0$, the probability density function has the form:

$$\lim_{\Delta_0 \to 0} P(\Delta, t; \Delta_0) \propto \frac{1}{t^{\frac{a+1-\xi}{2-\xi}}} \exp\left(-2\frac{\Delta^{2-\xi}}{(2b+c)(2-\xi)^2 t}\right) \tag{86}$$

Distribution of (86) represents a phenomenon known as spontaneous randomness. Such a phenomenon causes infinitesimally close Lagrangian trajectories to get separated in a finite time. Such a feature does not exist even in a chaotic regime with exponential separation.

Considering equation (63) in which $S(x, t)$ is a Gaussian random force acting on the system with correlation function $\langle S(x,t)S(x',t')\rangle = \delta(t-t')\chi(|x-x'|)$, (84) could be written as:

$$\frac{d}{dt}\langle f_t(\Delta)\rangle = \langle \mathcal{M}f_t(\Delta)\rangle + \chi(\Delta) \tag{87}$$

Here, by assuming $f(\Delta)$ as a two-point correlation function, $f(\Delta)$ satisfies following relation:

$$\mathcal{M} f_t(0) = \lim_{D \to 0} 2D \langle (\nabla \Phi)^2 \rangle \tag{88}$$

Now due to the spontaneous randomness in Lagrangian trajectories, (88) would be non-zero when $D \to 0$ and leads to scalar dissipative anomaly.

One can further investigates anomalies by calculating second-order structure function for the scalar:

$$S_2 = \langle \big(\Phi(t, \boldsymbol{x}) - \Phi(t, \boldsymbol{0})\big)^2 \rangle = 2 \langle f_t(0) - f_t(\Delta) \rangle$$

Assuming absorbing boundary condition on $\mathcal{M}$ (i.e. $f(0) = 0$), it can be shown that $S_2$ exhibits the anomalous scaling $\propto \Delta^{1-a}$ at its stationary form [55].

### 4.2.2 Path Integral Formulation

Here, a path integral formulation for a system with Langevin dynamics would be presented. The results derived here would be close to their quantum counterparts. The main difference would be the existence of complex numbers and complex fields in QFT whereas complex numbers do not play a structural role in classical theory. Lack of such numbers provide an opportunity to consider a path integral as a Wiener integral. This part is mostly based on the work conducted by Martin, Siggia, and Rose [56].

Consider the following Langevin equation:

$$\frac{\partial}{\partial t}\varphi(t,x) = -\Gamma \frac{\partial H(\varphi(t,x))}{\partial \varphi(t,x)} + f(t,x) \tag{89}$$

where $f(t)$ is random force with correlation function of:

$$\langle f(t,x)f(t',x')\rangle = 2D(|x-x'|)\delta(t-t') \tag{90}$$

a two-point correlation function could be written as a functional integral:

$$\langle \varphi(t_1,x_1)\varphi(t_2,x_2)\rangle = \langle \int D\varphi(t,x)\varphi(t_1,x_1)\varphi(t_2,x_2)\delta(\varphi(t,x) = a\ solution\ to\ (89))|f(t,x)\rangle \tag{91}$$

where R.H.S of (91) is the average over the realizations of random force.

The delta function in (91) could be rewritten to get:

$$\delta(\varphi(t,x) = a\ solution\ to\ (89)) = \delta\left(\frac{\partial}{\partial t}\varphi(t,x) + \Gamma \frac{\partial H(\varphi(t,x))}{\partial \varphi(t,x)} - f(t,x)\right) \tag{92}$$

Equation (92) could be then expressed as a functional integral of the form:

$$\delta\left(\frac{\partial}{\partial t}\varphi(t,x) + \Gamma \frac{\partial H(\varphi(t,x))}{\partial \varphi(t,x)} - f(t,x)\right) = \int D\tilde{\varphi}(t,x) \exp\left(-\int \tilde{\varphi}\left(\frac{\partial}{\partial t}\varphi(t,x) + \Gamma \frac{\partial H(\varphi(t,x))}{\partial \varphi(t,x)} - f(t,x)\right)dt\right) \tag{93}$$

substituting (93) in (91) would lead to:

$$\langle \varphi(t_1,x_1)\varphi(t_2,x_2)\rangle = \int D\varphi(t,x)D\tilde{\varphi}(t,x)\,\varphi(t_1,x_1)\varphi(t_2,x_2)\,e^{-S[\varphi,\tilde{\varphi}]} \tag{94}$$

where the action is written as:

$$S[\varphi,\tilde{\varphi}] = \int \frac{1}{2}\tilde{\varphi}D\tilde{\varphi} - \tilde{\varphi}\left(\frac{\partial}{\partial t}\varphi(t,x) + \Gamma \frac{\partial H(\varphi)}{\partial \varphi}\right)dt\,dx \tag{95}$$

and $\tilde{\varphi}$ is known as Martin-Siggia-Rose response field. For simplification, in the rest of the text, $(\varphi, \tilde{\varphi})$ would be denoted by $\varphi$.

Now, generating functional would be defined as:

$$\mathcal{G}[\mathcal{A}] = \int D\varphi \, e^{-S[\varphi] + \mathcal{A}\varphi} \tag{96}$$

where $\mathcal{A} \coloneqq (\mathcal{A}_\varphi, \mathcal{A}_{\tilde{\varphi}})$ and $\mathcal{A}\varphi \coloneqq \int \mathcal{A}_\varphi(x,t)\varphi(x,t) + \mathcal{A}_{\tilde{\varphi}}(x,t)\tilde{\varphi}(x,t) \, dt \, dx$.

Similar to its quantum counterpart, a n-point correlation function (Green's function) could be defined as:

$$\mathfrak{G}_n(x_1, t_1, \ldots, x_n, t_n) = \langle \varphi(t_1, x_1) \ldots \varphi(x_n, t_n) \rangle = \left. \frac{\delta^n \mathcal{G}[\mathcal{A}]}{\delta \mathcal{A}_\varphi(x_1, t_1) \ldots \delta \mathcal{A}_\varphi(x_n, t_n)} \right|_{\mathcal{A}=0} \tag{97}$$

The developed formulation could be used for perturbative methods where non-linear terms of action would be treated as perturbations. Decompose the term related to Hamiltonian in (89):

$$-\Gamma \frac{\partial H(\varphi(t,x))}{\partial \varphi(t,x)} = L\varphi(t,x) + V(\varphi(t,x))$$

where the first and second terms on R.H.S represent linear and non-linear terms of left-hand side.

By doing so, the action could be written as:

$$S[\varphi_1] = \int \frac{1}{2}\varphi_2 D \varphi_2 + \varphi_2 \left( -\frac{\partial}{\partial t}\varphi_1 + L\varphi_1 + V(\varphi_1) \right) dt \, dx \tag{98}$$

where $\varphi_2$ is the Martin-Siggia-Rose response field.

Subsequently, by writing the linear term of (98) in quadratic form of:

$$S[\varphi_1] = S_0[\varphi_1] + S_{int}[\varphi_1] = -\frac{1}{2}\int \begin{pmatrix}\varphi_1\\\varphi_2\end{pmatrix}^T \begin{pmatrix} 0 & \left(\frac{\partial}{\partial t}-L\right)^T \\ \frac{\partial}{\partial t}-L & -D \end{pmatrix}\begin{pmatrix}\varphi_1\\\varphi_2\end{pmatrix} dt\, dx +$$

$$\int \varphi_2 V(\varphi_1)\, dt\, dx \tag{99}$$

Dirac's notation could be employed for $S_0[\varphi_1]$ to have:

$$S_0[\varphi_1] = -\frac{1}{2}\int \langle \varphi|K|\varphi\rangle dt\, dx \tag{100}$$

where $|\varphi\rangle = \begin{pmatrix}\varphi_1\\\varphi_2\end{pmatrix}$ and $K = \begin{pmatrix} 0 & \left(\frac{\partial}{\partial t}-L\right)^T \\ \frac{\partial}{\partial t}-L & -D \end{pmatrix}$. Propagators associated with the first order of perturbation would be related to inverse of the matrix $K$.

$$\Delta := K^{-1} = \begin{pmatrix} \left(\frac{\partial}{\partial t}-L\right)^{-1} D \left(\left(\frac{\partial}{\partial t}-L\right)^T\right)^{-1} & \left(\frac{\partial}{\partial t}-L\right)^{-1} \\ \left(\left(\frac{\partial}{\partial t}-L\right)^T\right)^{-1} & 0 \end{pmatrix} \tag{101}$$

### 4.2.3. Mean Field Theoretic Approach to Scalar Advection

In this section, the formulation developed in 4.2.2 would be used for equations of scalar advection. In addition, renormalization techniques would also be introduced to calculate structure functions. This part is mostly based on the work reported by Hnatič et al [57,58] and Antonov [59]. Supplementary Information (Appendix C) could also be used for a brief introduction on QFT. Consider the generalized Kraichnan model where the velocity would satisfy a linear SDE with finite time correlation:

$$\frac{\partial}{\partial t}\boldsymbol{u} + \mathcal{L}\boldsymbol{u} = f \tag{102}$$

where $\mathcal{L} := \mathcal{L}(x)$ is a linear operator and $f$ is a random force representing the correlation function of:

$$\langle f_i(x,t)f_j(x',t')\rangle = D_{ij}^f(x,t;x',t') = \frac{1}{(2\pi)^{d+1}} \int d\omega \int d^d k\, P_{ij}(k)D^f(\omega,k)\, e^{-i\omega(t-t')}e^{ik.(x-x')} \tag{103}$$

Here, by assuming that $P_{ij}(k) = \left(\delta_{ij} - \frac{k_i k_j}{k^2}\right)$ is the projection tensor and $k = |k|$, and random force is white in time, (103) simplifies to:

$$D_{ij}^f(x,t;x',t') = \frac{D_0}{(2\pi)^d}\delta(t-t')\int d^d k\, P_{ij}(k)\, e^{-ik.(x-x')} \tag{104}$$

One can now couple (63) and (102) in the absence of a source to write an action functional of the form [58]:

$$S[\Phi] = \int -\frac{1}{2}u(D^u)^{-1}u + \frac{1}{2}\tilde{\Phi}D^\Phi\tilde{\Phi} - \tilde{\Phi}\left(\frac{\partial}{\partial t}\Phi + (u.\nabla)\Phi - D\nabla^2\Phi\right) dt\, dx \tag{105}$$

in which $D^u, D^\Phi$ are the correlators of the form:

$$D^u(x,t) = \langle u_i(x,t)u_j(0,0)\rangle = \frac{1}{(2\pi)^{d+1}}\int d\omega_k \int d^d k\, D^u(\omega_k,k)\, e^{-i\omega}\, e^{ik.x} \tag{106}$$

And a set of similar equations would apply for $D^\Phi$.

Furthermore, (105) could also be used for diagrammatic methods in Fourier space:

$$\langle \tilde{\Phi}\Phi\rangle = \frac{1}{-i\omega + Dk^2} = \langle \Phi\tilde{\Phi}\rangle^* \equiv \quad \Phi\,\text{-------}\blacktriangleright\text{-------}\,\tilde{\phi}$$

$$\langle u_i u_j\rangle = P_{ij}(k)D^u(\omega_k,k) \equiv \quad u_i\overline{\phantom{xx}}u_j$$

$$\langle \tilde{\Phi}u_j \frac{\partial}{\partial x^j}\Phi\rangle \equiv \quad u_j\overline{\phantom{xxxxxxxx}}\diagdown\,\begin{array}{c}\Phi\\ \tilde{\Phi}\end{array}$$

This could provide a proper setup for employing perturbation theory. However, such an iterative method could lead to certain poles in Ultra-Violet (UV) spectrum. This is a well-known problem and has been among the most important challenges hindering the introduction of a QFT of finite predictions. Such an obstacle was addressed by the methods of renormalization theory of Wilson [60]. In QED, renormalization has been used to solve scattering problems of all described fields in the standard model. In a classical problem, such as turbulence or phase transition, renormalization is necessary to understand large-scale behavior of the system and calculating correlations or structure functions of different variables.

For further development, a UV cut-off $\mu$ should first be set which is related to equation (105). Assume the relation $\frac{D_0}{D} = \zeta_0 = \mu^\varepsilon$ where $\varepsilon$ is a dimensionless parameter which its poles represent UV-divergences. Expressing the action functional (105) as $S[\Phi, D]$ in which $D$ is the only free parameter in (105), $S_R[\Phi, \mathfrak{D}, \mu]$ could be denoted as the renormalized functional which admits to UV-finite predictions. Here, $\mathfrak{D}$ represents renormalized parameter. Finally, renormalized action can be written as:

$$S_R[\Phi] = \int -\frac{1}{2}\boldsymbol{u}(D^u)^{-1}\boldsymbol{u} + \frac{1}{2}\tilde{\Phi}D^\Phi\Phi - \tilde{\Phi}\left(\frac{\partial}{\partial t}\Phi + (\boldsymbol{u}.\nabla)\Phi - \mathfrak{D}Z_\mathfrak{D}\nabla^2\Phi\right) dt\, dx \tag{107}$$

where $D = \mathfrak{D}Z_\mathfrak{D}$, $\zeta_0 = \zeta\mu^\varepsilon Z_\zeta$ and $Z_\zeta = Z_\mathfrak{D}^{-1}$.

Here $\mathfrak{D}$ and $\zeta$ are renormalized equivalents of $D$ and $\zeta_0$, and $Z_\mathfrak{D}$ is an independent renormalization constant.

A renormalized Green's function would then be calculated through:

$$\begin{cases} \mathfrak{G}_N^R(x_1, t_1, \ldots, x_n, t_n) = \frac{\delta^n \mathcal{G}^R[\mathcal{A}]}{\delta\mathcal{A}_\Phi(x_1,t_1)\ldots\delta\mathcal{A}_\Phi(x_n,t_n)}\bigg|_{\mathcal{A}=0} \\ \mathcal{G}^R[\mathcal{A}] = \int D\Phi\, e^{-S_R[\Phi] + \mathcal{A}\Phi} \end{cases} \tag{108}$$

For a transitionally invariant theory, one could with work connected parts of $\mathfrak{G}_n$, $\mathcal{W}_n$, and one-particle irreducible functions $\Gamma_n$. The generating functional for $\Gamma_n$ is defined as:

$$\Gamma[\Phi] = \mathcal{W}[\mathcal{A}] - \mathcal{A}\Phi \tag{109}$$

where

$$\mathcal{W}[\mathcal{A}] = \ln \mathcal{G}[\mathcal{A}] \tag{110}$$

For a renormalized theory, there should not be any UV-divergent poles in $\varepsilon$. A successful renormalized theory needs to be cut-off invariant. This implies that a renormalized connected part (i.e., $\mathcal{W}_n^R$) admits to the following equation:

$$\frac{d\mathcal{W}_n^R}{d\mu}(\mathfrak{D}, \mu, \varepsilon, \dots) = 0 \tag{111}$$

Accordingly, it could be concluded that:

$$\left(\mu\frac{\partial}{\partial \mu} + \Sigma_\zeta \mu \frac{\partial \zeta}{\partial \mu}\frac{\partial}{\partial \zeta} - \mu \frac{\partial \ln Z_\mathfrak{D}}{\partial \mu}\frac{\partial}{\partial \mathfrak{D}}\right)\mathcal{W}_n^R = 0 \tag{112}$$

where the anomalous dimension ($\gamma_\mathfrak{D}$) and beta function ($\beta_\zeta$) would be defined as:

$$\gamma_\mathfrak{D} = \mu \frac{\partial \ln Z_\mathfrak{D}}{\partial \mu} \tag{113}$$

$$\beta_\zeta = \mu \frac{\partial \zeta}{\partial \mu} = \zeta(-\varepsilon + \gamma_\mathfrak{D}) \tag{114}$$

In (112), $Z_\mathfrak{D}$ should be determined such that $\langle \tilde{\Phi}\Phi \rangle$ remains finite in the limit of $\varepsilon \to 0$. Through some calculations, it has been shown that [59]:

$$Z_\mathfrak{D} = 1 - \frac{\zeta(d-1)C_d}{2d\varepsilon} \tag{115}$$

$$\gamma_\mathfrak{D} = \frac{\zeta(d-1)C_d}{2d} \tag{116}$$

$$\beta_\zeta = \zeta \left(-\varepsilon + \frac{\zeta(d-1)C_d}{2d}\right) \qquad (117)$$

where $C_d = \frac{S_d}{(2\pi)^d}$ and $S_d$ is surface area of $d$-dimensional unit sphere.

From (117), Infra-Red (IR) attractive fixed point could be determined as:

$$\zeta^* = \frac{2d}{(d-1)C_d} \qquad (118)$$

By this definition, it could be shown that:

$$\beta_{\zeta^*} = 0 \ , \ \left.\frac{\partial \beta_\zeta}{\partial \zeta}\right|_{\zeta^*} = \varepsilon, \text{ and } \gamma_{\mathfrak{D}}^* = \varepsilon.$$

Now, the structure functions could be determined based on the fact that (112) must be satisfied.

For Kraichnan model, values of odd structure functions are zero. However, for even structure functions, with some dimensional considerations, it could be shown [59] that for inertial range defined as $\frac{1}{\mu} \ll |x - x'| \ll \frac{1}{m}$:

$$S_{2n} = \langle (\Phi(t,x) - \Phi(t',x'))^{2n} \rangle = \mathfrak{D}^n |x-x'|^{2n} R_{2n}\left(\frac{|t-t'|\mathfrak{D}}{|x-x'|^2}, \zeta, m|x-x'|\right) \qquad (119)$$

In large scale limit, coupling the constants would result in the IR attractive fixed point. Thus, by using (115) to (117), (119) becomes:

$$S_{2n} = \left(\frac{D}{\zeta^*}\right)^n |x-x'|^{(2-\varepsilon)n} Q_{2n}\left(\frac{|t-t'|D}{|x-x'|^{2-\gamma_{\mathfrak{D}}^*}}, m|x-x'|\right) \qquad (120)$$

In this equation, $Q_{2n}$ cannot be determined using (112). By analogy with the theory of critical phenomena, operator product expansion could be used to study $Q_{2n}(m|x-x'|)$ in inertial range. Interestingly, it should be noted that when $\varepsilon = \frac{4}{3}$, (120) reproduces results of Kolmogorov theory.

## 4.3. Intermittency

Intermittency was mentioned when discussing anomalous scaling in Burgulence. In this section, a more detailed discussion of intermittency would be presented.

Recall the scaling (53) could lead to a scale invariance for structure functions, meaning that:

$$S_n(\Delta x, t) \propto (|\Delta x|)^{\frac{n}{3}} \tag{121}$$

Such a relation, which would be a result of Kolmogorov theory, indicates that the generalized kurtosis of velocity profiles remains constant:

$$\frac{S_{2n}(\Delta x,t)}{S_n(\Delta x,t)} \propto const \tag{122}$$

However, it has been shown that the experimental data would not be in agreement with (122). For instance, $S_6(\Delta x, t) \propto (|\Delta x|)^{\xi_6}$ where $\xi_6$ deviates from 2 and most likely has a value of almost 1.78 which would be inconsistent with Kolmogorov's universality hypothesis. Such an inconsistency is evidently a violation of Kolmogorov's theory and is due to intermittency.

There has been many models of intermittency which would not be discussed here and readers are referred to detailed explanations of such models (e.g., $\beta$ model or bifractal model) provided in literature [38]. Herein, the multifractal model proposed by Parisi and Frisch [61] would be introduced and after discussing its basic properties, general features of multifractals and intermittency would be assessed in the context of stochastic analysis. Eventually, based on SDEs, an alternative approach for intermittency would also be mentioned.

### 4.3.1. Multifractal Model of Intermittency

Consider Euler equation:

$$\frac{\partial u}{\partial t} + (u \cdot \nabla)u = -\nabla p \tag{123}$$

This equation is invariant under transformation:

$$x \to \lambda x \quad u \to \lambda^h u \quad t \to \lambda^{1-h} t$$

In (123) the absence of viscosity implies the existence of infinitely many scaling parameters $h$. Using this intuition and considering high Reynolds limit, one could diminish the principle of existence of a global scale invariance factor $h$ to many local scale invariance factors.

Define:

$$\delta u(\Delta x, x) := |u(x + \Delta x, t) - u(x, t)|$$

Based on Kolmogorov's theory $\delta u(\Delta x, x) \propto |\Delta x|^{\frac{1}{3}}$. However, the multifractal model would be based on the following assumption:

*if $x \in S_h$ then there exists one and only one $h \in [h_m, h_M]$ such that $\delta u(\Delta x, x) \propto |\Delta x|^h$*

where $S_h$ is a fractal set with Hausdorff dimension $D_H(h)$. Here, $h_m, h_M$ would also be postulated to be universal.

Furthermore, it could be deduced from the volumetric argument that:

$$P(\Delta x, h) \propto |\Delta x|^{d-D_H(h)} \tag{124}$$

and

$$S_n(\Delta x, t) \propto \int P(\Delta x, h) \, \delta u(\Delta x, x) \, d\mu(h) = \int |\Delta x|^{d-D_H(h)+nh} \, d\mu(h) \tag{125}$$

where $\mu(h)$ is the Lebesgue measure of fractal set $S_h$.

To calculate $\xi_n = \lim_{|\Delta x| \to 0} \frac{\ln n(\Delta x, t)}{\ln |\Delta x|}$, the steepest decent method could be employed, resulting in:

$$\xi_n = \inf_h (nh + d - D(h)) \tag{126}$$

Notice (126) could be seen as a Legendre transformation which provides a geometrical interpretation for $D(h)$.

So far, there are no strategies to calculate $D(h)$ from first principles and it is believed that such computation is not attainable at present [62]. However, there have been some models that have tried to approximate $D(h)$ without using of the first principles. One of such models is called random $\beta$ model developed by Benzi and colleagues [63]. To construct this model, the Richardson cascade was first partitioned (similar to shell models). It was assumed that a mother eddy in step q of cascade with characteristic size $l_q \sim \frac{L}{2^q}$ would split into daughter eddies that would cover the $\beta_q$ fraction of mother eddy volume. By the fact that the energy transfer is constant in the cascade, the following relation for velocity difference on scale $l_q$ holds:

$$\delta u_q(\Delta x, x) \propto (l_q)^{\frac{1}{3}} \prod_{j=1}^{q} (\beta_j)^{-\frac{1}{3}}$$

Therefore:

$$\xi_n = \frac{n}{3} - \log_2 \langle \beta^{1-\frac{n}{3}} \rangle \tag{127}$$

Here $\beta_j$ could be considered as a random variable following a binomial distribution as:

$$\begin{cases} P\left(\beta_j = \frac{1}{2}\right) = x \\ P(\beta_j = 1) = 1 - x \end{cases}$$

that consequently leads to:

$$\xi_n = \frac{n}{3} - \log_2\left(x + (1-x)2^{\frac{n}{3}-1}\right) \tag{128}$$

Now, $x$ could be calibrated in (128) $x$ to fit the experimental data. Furthermore, one could use this model to compute the probability density function for velocity increments [64].

### 4.3.2. Multifractals and Intermittency: A Mathematical Description

In this part, a general description of intermittency and multi-fractality would be studied based on the work of Khoshnevisan et al [65]. Some examples have been presented and references have been provided to the readers to better understand the theories and mathematical details behind such results.

Consider the parabolic Anderson model:

$$\frac{\partial}{\partial t}u(t,x) = \frac{1}{2}\frac{\partial^2}{\partial x^2}u(t,x) + u(t,x)\zeta(t,x) \tag{129}$$

where $\zeta(t,x)$ is a white noise with Schwartz type distribution. It is well established that not only (129) has a unique solution, but also the following equality holds for its $k^{\text{th}}$ moment Lyaponouv exponent:

$$\lambda_k = \lim_{t\to\infty} \frac{\ln\langle|u(t,x)|^k\rangle}{t} = \frac{k(k^2-1)}{24} \tag{130}$$

Consider the flatness of function $f(t)$:

$$F := \frac{\langle|u(t,x)|^4\rangle}{(\langle|u(t,x)|^2\rangle)^2} \tag{131}$$

Now, assuming the Fourier transform of $u(t,x)$ denoted as $\hat{f} = \int dt\, e^{-i\omega} u$, one can get:

$$f = \int d\omega\, e^{i\omega t}\, \hat{u}$$

$$f_\Omega := \int_{\omega > \Omega} d\omega\, e^{i\omega t} \hat{u}$$

$$F_\Omega := \frac{\langle |f_\Omega(t,x)|^4 \rangle}{(\langle |f_\Omega(t,x)|^2 \rangle)^2}$$

If $F_\Omega$ grows without bounds with respect to $\Omega$, then $f$ would be intermittent. From (130), as $\frac{\lambda_k}{k}$ is a strictly increasing function of $k$, it could be concluded that $u(t,x)$ would satisfy the constraints for intermittency.

Assume function $g(|t|)$ as an increasing function and that the following holds:

$$\limsup_{t \to \infty} \frac{|f(t)|}{g(|t|)} = 1 \quad \text{a.s.}$$

By defining the following random set:

$$\mathfrak{R}_{f,g}(\Lambda) = \left\{ t: t > c_0, \frac{|f(t)|}{g(|t|)} > \Lambda \right\}$$

where $\Lambda$ and $c_0$ are positive constants, multi-fractality could be defined as below:

Consider a family of infinite sets of ordered length scales $\mathcal{T} = \{\Lambda_i : 0 < \Lambda_1 < \Lambda_2 < \cdots \}$. Then $f$ would have a multifractal behavior in gauge $g$ if there exists a $\mathcal{T}$ such that $D_H\left(\mathfrak{R}_{f,g}(\Lambda_{i-1})\right) < D_H\left(\mathfrak{R}_{f,g}(\Lambda_i)\right)$.

Back to (129), it can be shown that $g(x) = (\ln_+ x)^{\frac{2}{3}}$ is a natural gauge for $X(t,x) := \left(\frac{32}{9t}\right)^{\frac{1}{3}} \ln u(t,x)$ with $D_H\left(\Re_{X,g}(\Lambda)\right) = 1 - \Lambda^{\frac{3}{2}}$. Moreover, since there are one-to-one correspondences between peaks of a function and its logarithm, it could be deduced that tall peaks of $u(t,x)$ would be multifractal.

Now, the question is whether multi-fractality and intermittency are equivalent and coincide in all cases or not. Khoshnevisan et al [65] have argued and confirmed that the answer to this question is no. They studied random heat equation with additive white noise in time and space and showed that, similar to that of (129), following equation also satisfies a unique solution, although the moment Lyapunov exponents of all orders would be equal to zero:

$$\frac{\partial}{\partial t} v(t,x) = \frac{1}{2}\frac{\partial^2}{\partial x^2} v(t,x) + \varphi(t,x) \tag{132}$$

This implies that the solution to (132) would not be intermittent.

In terms of the multi-fractality of $v(t,x)$, the following result could be obtained:

define $X(t,x) := \left(\frac{\pi}{t}\right)^{\frac{1}{4}} v(t,x)$, $g(t) = (\ln_+ t)^{\frac{1}{2}}$. The peaks of $X(t,x)$, and consequently $v(t,x)$, would be multifractal in gauge $g(t)$ for all $t > 0$ and $D_H\left(\Re_{X,g}(\Lambda)\right) = 1 - \Lambda^2$.

### 4.3.3. Stochastic Hierarchical Modeling of Intermittency

As mentioned before, the multifractal model could be used to calculate the probability for velocity increments. In this section, an alternative for such calculation would be discussed. Most of the provided discussions are based on the work of Salazar and Vasconcelos [66].

Similar to shell models, a partitioning on the energy cascade could be carried out. In the $n^{th}$ interval of this partition, the energy transfer rate would be denoted by $\epsilon_n = \epsilon(r_n)$ with the scale $r_n = \frac{L}{2^n}$.

To model the dynamics of $\epsilon_n$, the following hierarchical set of SDEs could be used:

$$d\epsilon_i = \alpha_i(\epsilon_{i-1} - \epsilon_i)dt + \beta_i \epsilon_i \, dW_i \tag{133}$$

in which $\alpha_i$ and $\beta_i$ are time-independent constants. Such an equation could be used to reproduce the results of Kolmogorov's theory on average since $\langle \epsilon_i \rangle = \epsilon_0$. Considering that partitioning would be applied to energy cascade, existence of a multiplicative noise seems rather natural. Exploiting the fact that time scales would also be separated in different steps of cascade, it could be assumed that $\epsilon_i$ would change in constant $\epsilon_{i-1}$. Using such assumption, one can derive the forward Kolmogorov equation of (133):

$$\frac{d}{dt} P(\epsilon_i | \epsilon_{i-1}) = -\frac{\partial}{\partial \epsilon_i}\left(\alpha_i(\epsilon_{i-1} - \epsilon_i) P(\epsilon_i | \epsilon_{i-1})\right) + \frac{\partial^2}{\partial \epsilon_i^2}\left((\beta_i \epsilon_i)^2 P(\epsilon_i | \epsilon_{i-1})\right) \tag{134}$$

In stationary state, it would be simple to solve (134) to achieve:

$$P(\epsilon_i | \epsilon_{i-1}) = \frac{(\kappa_i \epsilon_{i-1})^{\kappa_i+1}}{\Gamma(\kappa_i+1)} \epsilon_i^{\kappa_i-2} e^{\frac{\kappa_i \epsilon_{i-1}}{\epsilon_i}} \tag{135}$$

where $\kappa_i = \frac{2\alpha_i}{\beta_i^2}$ and $\Gamma(z+1) = \int_0^\infty t^z e^{-t} dt$. On the other hand:

$$P(\epsilon_i) = \int P(\epsilon_i | \epsilon_{i-1}) P(\epsilon_{i-1}) \, d\epsilon_{i-1} = \int \ldots \int \prod_{j=1}^{i-1} P(\epsilon_j | \epsilon_{j-1}) \, d\epsilon_{i-1} \ldots d\epsilon_1 \tag{136}$$

By assuming homogeneity of velocity increments, $\delta \boldsymbol{u}(r, \boldsymbol{x})$ would only be just a function of $r$ (i.e., $\delta \boldsymbol{u}(r, \boldsymbol{x}) = \delta \boldsymbol{u}(r)$). Now, (136) could be used to calculate the distribution of velocity increments $\delta \boldsymbol{u}(r)$ as:

$$P(\delta \boldsymbol{u}(r)) = \int P(\delta \boldsymbol{u}(r) | \epsilon_r) P(\epsilon_r) \, d\epsilon_r \tag{137}$$

Considering a Gaussian from for $P(\delta u(r)|\epsilon_r)$, (137) could be expressed in terms of generalized hypergeometric functions of order $(n, 0)$ and $F_n^0$:

$$P(\delta \boldsymbol{u}(r)) = \frac{1}{\sqrt{2\pi}} \prod_{i=1}^{n} \left( \frac{\Gamma\left(\kappa_i + \frac{3}{2}\right)}{\sqrt{\kappa_i} \Gamma(\kappa_i + 1)} \right) F_n^0 \left( \kappa_1 + \frac{3}{2}, \dots, \kappa_n + \frac{3}{2}; -\frac{(\delta \boldsymbol{u}(r))^2}{2\kappa_1 \dots \kappa_n} \right) \tag{138}$$

By calibrating $n$ and $\kappa_i$ (usually $\kappa_i = \kappa \ \forall i$), the experimental data could now be fitted. Such an operation for low temperature helium jet has already been conducted [66].

## 4.4 Incompressible Turbulence

Incompressible turbulence is referred to turbulent behavior of a system which its dynamics obey (1). So far, much simpler problems such as Kraichnan model or Burgers turbulence have been tackled. Such tackling have given away a few ideas about what it means to actually solve turbulence. Such a goal proved to be an extremely difficult task even for simple models and with our current knowledge, it is a formidable task to confront incompressible turbulence.

Although progresses have been made on two-dimensional turbulence, still serious obstacles remain in the way of solving three-dimensional turbulence. Obstacles such as the lack of a uniqueness theorem, ergodic theorems for strong solutions, difficulty of numerical solutions even with current computation power, intermittency, and certain anomalies like dissipative anomaly. In two-dimensions, besides less complexity in numerical simulation, progresses in generating realistic experimental information through methods such as Electromagnetically Forced Conducting Fluid Layers or Soap-Film Channels [67] have provided the scientists with appropriate data to delve into two dimensional scenarios. However, although promising results have been achieved, it should be noted that even two-dimensional turbulence is still far from being fully solved. It should be noted that two-dimensional models that have been perturbed by three-dimensional noises have already been used to model and simulate geophysical flows, meteorological events, and some other environmental and industrial process.

In the following sections, first a more exact comparison between two and three-dimensional cases would be provided, and then two-dimensional turbulence would be discussed in more details and some of the recent advances on the topic would be covered.

### 4.4.1. Reduction in Dimensionality

Taking the curl of (1), the following equation holds:

$$\begin{cases} \frac{\partial \boldsymbol{\omega}}{\partial t} + (\boldsymbol{u} \cdot \nabla)\boldsymbol{\omega} = \nu \nabla^2 \boldsymbol{\omega} + (\boldsymbol{\omega} \cdot \nabla)\boldsymbol{u} \\ \boldsymbol{\omega} = \nabla \times \boldsymbol{u} \end{cases} \quad (139)$$

where $(\boldsymbol{\omega} \cdot \nabla)\boldsymbol{u}$ is known as vortex stretching that in two-dimensions is equal to zero. Therefore, in two-dimensions:

$$\frac{\partial \boldsymbol{\omega}}{\partial t} + (\boldsymbol{u} \cdot \nabla)\boldsymbol{\omega} = \nu \nabla^2 \boldsymbol{\omega} \quad (140)$$

As shown, although (140) resembles (63), it should be considered that $\boldsymbol{\omega}$ is strongly dependent on $\boldsymbol{u}$, whereas this is not necessarily the case in (63). Furthermore, writing equations for energy of the system $E$ would lead to:

$$E = \langle \tfrac{1}{2} \boldsymbol{u}^2 \rangle$$

$$\frac{d}{dt} E = -2\nu \langle \tfrac{1}{2} \boldsymbol{\omega}^2 \rangle = -2\nu \Omega \quad (141)$$

where $\Omega$ is called enstrophy and obeys:

$$\frac{d}{dt}\Omega = -2\nu \langle \frac{1}{2}(\nabla \omega)^2 \rangle \tag{142}$$

Notice that (141) and (142) indicate that both enstrophy and energy are constants of motion in the Euler equation. More interestingly, enstrophy is a decreasing function, and thus, is bounded to its initial value. This implies that no energy dissipative anomaly exists. However, in these cases, another form of anomaly should be considered. In the inviscid limit, (142) becomes independent of viscosity in stationary state and thereby, enstrophy dissipation anomaly exists in two dimensions. In addition, combination of (142) and (141) indicates that energy would be almost conserved in the limit of zero viscosity. The energy in Fourier space could be represented as:

$$E = \int E(k)dk \tag{143}$$

$$\Omega = \int k^2 E(k)dk \tag{144}$$

Since energy is conserved, the surface area under $E(k)$ would be constant while the surface area under $k^2 E(k)$ would shrink. A scenario that satisfies these conditions could be the growth of $E(k)$ at smaller $k$ while being depleted at larger $k$ values. This interpretation is quite interesting as it points out to an inverse cascade of energy from lower to larger scales. The idea of existence of an inverse cascade in two dimensions was first introduced by Kraichnan [68]. He also proposed the existence of a direct enstrophy cascade. Today, the existence of such cascades has been well-established. For detailed discussion of experimental results on these cascades one could be referred to [67,69].

With dimensional analysis, it can be shown that $E(k) \propto k^{-\frac{5}{3}}$ for inverse energy cascade regime and $E(k) \propto k^{-3}$ for direct enstrophy cascade regime. On the other hand, $k^{-3}$ could cause some contradictions. Assume a series of wavenumbers in enstrophy cascade regime $k_0, k_1, \ldots$ where $k_i = 0.1 k_{i-1}$. Accordingly:

$$\Omega_{DC} \propto \int k^2 E(k) dk = \int_{k_0}^{k_1} k^{-1} + \int_{k_1}^{k_2} k^{-1} + \cdots = \sum_i \ln(0.1)$$

where $\Omega_{DC}$ is the contribution of direct cascade regime to total enstrophy. It can be seen that different length scales contribute equally to $\Omega_{DC}$, confirming the non-locality of energy spectrum. However, existence of cascade relies on a locality assumption in distribution of energy. Therefore, the result and the assumption contradict each other.

There has been some experimental and numerical investigations on the possibility of intermittency in inverse cascade [70,71] in which the results, in contrast to the three-dimensions, did not show any considerable intermittency. However, it should be noted that this result does not rule out the existence of intermittency in two-dimensions

Turbulence, both in two and three-dimensions, is far from being completely disorganized. Anyone who looks at the unforced motion of smoke in still air, can recognize swirls that form instantaneously and sometimes evolve to considerable sizes. Same patterns fascinated Leonardo da Vinci when he first studied turbulence. Such coherent structures form due to some instabilities such as Kelvin-Helmholtz instability. However, being moved in three dimensions, such structures would be more notable than those in two-dimensional flows. In turbulent flow, small patches of vorticity would go through a series of filamentations and eventually form some lasting structures, vortex filaments, with their own peculiar dynamics. These filaments could then merge or break up and disappear, feeding the cascades during their evolution. As they grow in size, the inverse cascade would be supplied. On the other hand, they admit to complex structures which cause sharp vorticity and velocity gradients which could affect the spectrum throughout all its domain.

Being inspired by such structure, Lars Onsager employed equilibrium statistical mechanics for his model of point vortices. His approach was quite interesting and the most astonishing results of his theory was the prediction of absolute negative temperature states in sufficiently high energies. Such a phenomenon takes place when the system has an upper energy bound that leads to a negative $\frac{\partial S}{\partial E}$ ($S$ is the entropy) [72]. This idea was later further developed by Montgomery and Joyce and an equation for large scale vortex solutions was derived [73].

Another interesting phenomenon hypothesized by Kraichnan and supported by experimental evidence [69] is the existence of condensate states [68]. When the dissipation range $\eta$ exceeds the system size, inverse cascade causes energy to move to larger scales. However, at some points, the energy dissipation would terminate the cascade. It would be expectable for states in a system to pile in smallest wave numbers. Such states take their name by analogy to Bose-Einstein condensation.

Many breakthroughs have been made by mathematicians and physicists when exploring two-dimensional turbulence. Ergodic theorems have been established, statistical theories have been developed, and the structure of two-dimensional flow have been investigated quite successfully. However, three-dimensional case, the highest branch of turbulence tree (Fig 1), have not practiced much improvements and have stood still since Kolmogorov. While easier to observe in experiments, three-dimensional problems are harder to simulate numerically. No global uniqueness or strong ergodic theorem has been established so far and even Kolmogorov's description has been challenged by intermittency. Coherent structures in three-dimensions are not as trivial as the two-dimensional case where Burgers vortices, vortex sheets, and vortex filaments are also probable candidates for coherent structures. All these characteristics have made three-dimensional turbulence a much more complicated problem to deal with.

### 4.4.2. Statistical Mechanics of Two-Dimensional Flow

In this section, the theory of Onsager and Montgomery would be first introduced [73]. Assume there exists $N$ point vortices in a plane. The coordinates of each vortex could then be represented by $\mathbf{r}_i = (x_i, y_i)$ with vorticity of $\Gamma_i$. Therefore, the Hamiltonian of system could be written as:

$$H = -\frac{1}{2\pi}\sum_{i<j} \Gamma_i \Gamma_j \ln(r_{ij}) \tag{145}$$

where $r_{ij} = |\mathbf{r}_i - \mathbf{r}_j|$. Hamilton equations have the form:

$$\Gamma_i \dot{x}_i = \frac{\partial H}{\partial y_i}, \quad -\Gamma_i \dot{y}_i = \frac{\partial H}{\partial x_i}, \quad \frac{dH}{dt} = 0 \tag{146}$$

The total circulation can be written in terms of vorticity species as following:

$$\omega_a = \Gamma_a n_a(\mathbf{r}, t), \quad \omega = \sum_a \omega_a \tag{147}$$

where $n_a$ is the density of species $a$. In addition, vorticity and stream function $\psi$, would satisfy the Poisson equation:

$$\nabla^2 \psi = \omega \tag{148}$$

To use the techniques of equilibrium statistical mechanics, existence of an equilibrium state would be essential. By assuming the existence of such states, one could reach the following expression for the entropy of the system [74]:

$$S[\omega_a] = -\sum_a \int \frac{\omega_a}{\Gamma_a} \ln(\frac{\omega_a}{\Gamma_a N_a}) \, d\boldsymbol{r} \tag{149}$$

where $N_a = \int n_a(\boldsymbol{r},t) \, d\boldsymbol{r}$.

Now, considering the conservation laws governing the system, one can maximize (149). In the case of no dissipation, energy and total vorticity would be conserved. Thus:

$$\delta S[\omega_a] - \beta \delta E - \sum_a \alpha_a \delta(\Gamma_a N_a) = 0 \tag{150}$$

where $\delta E = \int \psi \delta\omega \, d\boldsymbol{r}$ and $\delta \Gamma_a N_a = \int \delta \omega_a \, d\boldsymbol{r}$.

As a result:

$$\omega = \sum_a A_a \Gamma_a e^{-\beta \Gamma_a \psi} \tag{151}$$

in which $A_a = \frac{N_a}{\int e^{-\beta \Gamma_a \psi} \, d\boldsymbol{r}}$ and $\beta = \frac{1}{kT}$.

As an example, consider the case in which only two vorticity species with $\Gamma_\pm = \pm \Gamma$ and $N_\pm = \frac{N}{2}$ exist. Thus, the sinh-Poisson equation could be derived from (148) and (151) which would govern the equilibrium stream function as:

$$\nabla^2 \psi = 2\Gamma A \sinh(-\beta \Gamma \psi) \tag{152}$$

where $A = 2A_+ = \frac{1}{\langle e^{-\beta \Gamma \psi} \rangle} = 2A_- = \frac{1}{\langle e^{\beta \Gamma \psi} \rangle}$. In this equation, it has been assumed that $A_- = A_+$, which in the context of classic statistical mechanics, indicates that both species have the same chemical potential.

The second law of thermodynamics provides us with a Lyapunov function for the dynamics of any arbitrary system. Such a function could be used to predict the equilibrium state of the system. However, in the case of non-dissipative two-dimensional fluid motion, there exists a natural

Lyapunov function for the system, namely enstrophy. Accordingly, based on (142), a minimum enstrophy principle could be used to predict the equilibrium state of the system [75]. There have been cases such as strong mixing in which these two principals have shown to be equivalent to each other [74,76].

In deriving (150), existence of an equilibrium state was assumed. Validity of such assumption however, needs to be verified. Looking at point vortices in inviscid two-dimensional fluid flow as a Hamiltonian system extends beyond equations (145) and (146); There is a Liouville theorem for such a system. If $\omega = \sum_{i=1}^{N} \Gamma_i \delta(\boldsymbol{r} - \boldsymbol{r}_i)$ and by considering $\nabla^2 \psi = \omega$, $\omega$ could be expressed in terms of Laplacian eigenfunctions, $e_i(\boldsymbol{r})$: $\omega = \sum_{i=1}^{N} \gamma_i \, e_i(\boldsymbol{r})$. It can be shown that the following Liouville theorem holds [77]:

$$\frac{\partial \dot{\gamma}_i}{\partial \gamma_i} = 0 \quad \forall i$$

Such a theorem helps building a microcanonical measure:

$$\mu_{\mathcal{MC}} \propto \prod_{i=1}^{N} d\gamma_i \, \delta(E[\gamma_i] - E_i) \, \delta(\Gamma[\gamma_i] - \Gamma_i) \tag{153}$$

(153) is an invariant measure of system. Rather interestingly it can be shown that Euler equation corresponding to the inviscid case describes a non-ergodic system. Even so, like many other non-ergodic systems in equilibrium statistical mechanics, we can assume that ergodicity is not completely absent in the system, but rather perturbed to some small extent. There is no reason to suspect such assumption as predictions of statistical mechanics do agree with experiments to a significant scale. For the Navier-Stokes equation, ergodicity was discussed in section 3.2 and under some mild assumptions, ergodicity was ensured. For further information on statistical mechanics of two-dimensional turbulence, the study conducted by Delbaen could be considered [78].

### 4.4.3. Stochastic Dynamics of Two-dimensional Hydrodynamics

Spatiotemporal patterns in dynamics of coherent structure seem rather unpredictable and one of the main questions that arises is the existence of attractors in the system. First, recall the following definitions:

$(\Omega, \mathcal{F}, \mathbb{P})$ *is defined as the probability space*

$\theta^s: \Omega \to \Omega$ *as the shift oprator:* $\theta^s \pi(t) = \pi(t+s)$ *for* $\pi \in \Omega$

*A continuous random dynamical system over some separable Banach space X as the mapping* $\Phi: \mathbb{R}^+ \times \Omega \times X \to X$ *such that* $(t, \pi, x) \to \Phi(t, \pi, x)$ *for* $x \in X$ *with following conditions:*

*i)* $\Phi(0, \pi, x) = x$

*ii)* $\Phi(t + s, \pi, x) = \Phi(t, \theta^s \pi, \Phi(s, \pi, x))$

*iii)* $\Phi(t, \pi, .)$ *is continuous.*

*A random set* $\{A(\pi)\}_{\pi \in \Omega} \subset X$ *is said to be a* $\mathcal{D}$ *random attractor if:*

*i)* $A(\pi)$ *is compact*

*ii)* $\{A(\pi)\}_{\pi \in \Omega}$ *is invariant, i.e.,* $\Phi(t, \pi, A(\pi)) = A(\theta^t \pi)$

*iii) For a collection of random subsets of X, denoted by* $\mathcal{D} = \cup \{D(\pi)\}_{\pi \in \Omega}$:

$\lim_{t \to \infty} d(\Phi(t, \theta^{-t}\pi, D(\theta^{-t}\pi)), A(\pi)) = 0$ *for every set in* $\mathcal{D}$, *where* $d(A, B) = \sup_{x \in A} \inf_{y \in B} |x - y|_X$.

*Furthermore, defining an absorbing set* $\{R(\pi)\}_{\pi \in \Omega} \in \mathcal{D}$ *if there exisits some* $T(\pi)$ *such that:*

$\Phi(t, \theta^{-t}\pi, D(\theta^{-t}\pi)) \subset R(\pi)$ *for every set in* $\mathcal{D}$ *and every* $t \geq T(\pi)$

*Φ is called 𝔇 pullback asymptotically compact in X if the sequence $\{\Phi(t_n, \theta^{-t_n}\pi, x_n)\}_{n\geq 1}$ has a convergent subsequence when $t_n \to \infty$ and $x_n \in D(\theta^{-t_n}\pi)$.*

It could be shown that two-dimensional stochastic Navier-Stokes with an additive noise (26) possesses a unique random attractor [79]. This has been explained in detail and proved by Xiu and Karniadakis using the general result of SDEs which states that if the dynamical system corresponding to a SDE has an absorbing set $\{R(\pi)\}_{\pi \in \Omega}$ and the system is 𝔇 pullback asymptotically compact in X, then $A(\pi) = \cap_{T \geq 0} \overline{\cup_{t \geq T} \Phi(t, \theta^{-t}\pi, R(\theta^{-t}\pi))}$ is a unique 𝔇 random attractor.

The random dynamical system associated to (26) is defined as:

$$\Phi(t, \pi, u_0(\pi)) = u(t, \pi)$$

The strategy here is to prove $\Phi$ satisfies 𝔇 pullback asymptotically compactness and admitting to an absorbing set. After some analytical steps such as estimating norm of dynamical system in corresponding Hilbert space, the validity of such random dynamical system could be confirmed.

There has been some work on non-equilibrium thermodynamics and statistical mechanics of two-dimensional hydrodynamics using stochastic Navier-Stokes. In the recent work conducted by Wu and Wang [80], an interacting many body interpretation of Navier-Stokes was established in wavevector space. First writing stochastic Navier-Stokes in wavevector space:

$$\frac{\partial}{\partial t} u(\boldsymbol{k}, t) = -i P_{ij} \sum_{\boldsymbol{k}'+\boldsymbol{k}''=\boldsymbol{k}} k_m u_m(\boldsymbol{k}'') u_j(\boldsymbol{k}') - \nu k^2 u(\boldsymbol{k}, t) + f(\boldsymbol{k}, t) \quad (154)$$

where $P_{ij}(k) = \left(\delta_{ij} - \frac{k_i k_j}{k^2}\right)$ is the transverse projection operator, and:

$$\langle f_i(\boldsymbol{k}', t'), f_j(\boldsymbol{k}, t) \rangle = P_{ij}(k) D_0 \delta(t - t') = D_{ij}^u(k) \quad (155)$$

Or in the matrix form:

$$\langle f(\mathbf{k'}, t'), f(\mathbf{k}, t)\rangle = \widetilde{\mathbf{P}}(k) D_0 \delta(t - t') = \mathbf{D^u}(k)$$

Now, consider only the local terms, i.e., stochastic and friction force. These terms form a Langevin equation for Fourier modes, therefore using Einstein-Smoluchowski relation (Supplementary Information, Appendix A), an effective temperature expression could be derived for the system as:

$$T_{eff}(\mathbf{k}) = D_0 \frac{L^3}{2\nu k^2} \tag{156}$$

Therefore, one can think that each Fourier mode would be in contact with a heat bath of temperature (156) while there exists a non-local convective force (due to non-linearity) which causes different modes to interact.

Using Fokker-Planck equation, energy balance and entropy production for the system could be derived as:

$$\frac{d}{dt} E(\mathbf{k}, t) = T(\mathbf{k}, t) - \dot{Q}(\mathbf{k}, t) \tag{157}$$

where:

$$\Phi(\mathbf{k}, t) = -\mathbf{D^u}(k) \cdot \nabla_{u(-k,t)} \ln\left(\frac{P_t(u)}{P_0(u)}\right)$$

$$\dot{Q}(\mathbf{k}, t) = \langle \sum_k \Phi(\mathbf{k}, t) \cdot \nabla_{u(k,t)} \sum_{k'} u(\mathbf{k'}, t) \cdot u(-\mathbf{k'}, t) \rangle$$

$$T(\mathbf{k}, t) = Re\left(L^3 \langle -iP_{ij} \sum_{k'+k''=k} k_m u_m(\mathbf{k''}, t) u_j(\mathbf{k'}, t) \rangle\right)$$

Here $T(\mathbf{k}, t)$ is the energy transfer associated with convective force and $\dot{Q}(\mathbf{k}, t)$ is heat flow rate from Fourier modes to heat baths. Moreover, an expression for the non-equilibrium entropy could also be derived:

$$S(t) = -\int P_t(u) \ln(P_t(u)) \, \delta u \tag{158}$$

From (158), one can show that the entropy balance equation has the form of:

$$\frac{d}{dt} S(t) = \sum_k \frac{\dot{Q}(k,t)}{T_{eff}(k)} + \langle \Phi(-\mathbf{k}, t) D^{u-1}(k) \Phi(\mathbf{k}, t) \rangle \tag{159}$$

where the first term in R.H.S of (159) is associated with total entropy flow and second term is related to entropy production.

Through another approach considered by Bouchet and colleagues [81,82], based on Onsager–Machlup formalism, fluctuation paths starting from an attractor of two-dimensional Euler system was studied. For a general Langevin equation:

$$dX(t) = a(X)dt + b(X)\, dW_t$$

Discretization could be then considered to get:

$$X(t_{n+1}) = X(t_n) + a(X(t_n))\Delta t + b(X(t_n))\sigma_i \sqrt{\Delta t} \tag{160}$$

where:

$t_k \in \{t_n\}_{n \geq 0}$ such that $t_{k+1} - t_k = \Delta t$ and

$$dP = \exp(\sum_i \sigma_i^2) \prod_i \frac{d\sigma_i}{\sqrt{2\pi}} \tag{161}$$

Inverting (161), substituting it in (160), and taking the limit $\Delta t \to 0$ would result in a functional formulation of probability measure of a particular trajectory as:

$$dP(x_1, x_2, \ldots, x_n) = D[x] J[\sigma|x] \exp\left(-\frac{1}{2}\int L(x, \dot{x}, t) \, dt\right) \quad (162)$$

and

$$L(x, \dot{x}, t) = \frac{1}{2}\left(\frac{\dot{x}-a(x)}{b(x)}\right)^2$$

Accordingly, the action could be defined as:

$$\mathcal{A}[x_0, x_T, T] = \int_0^T \frac{1}{2} L(x, \dot{x}, t) \, dt \quad (163)$$

Now, it would be possible to calculate quantities such as the most probable path from an attractor to any arbitrary point. By assume $x_a$ belongs to some attractor, the most probable trajectory from $x_a$ to an arbitrary point $x_b$ would be a minimizer of the action (163):

$$\mathcal{A}[x_a, x_b, \infty] = \min_x\{\mathcal{A}[x]: \lim_{T \to \infty} x(-T) = x_a, x(0) = x_b\} \quad (164)$$

As discussed, Euler dynamics is non-ergodic, it would be possible that more than one attractor exist in the system. In such case, instantons could be defined as:

$$\mathcal{A}[x_a, x_b] = \min_x\{\mathcal{A}[x]: \lim_{T \to \infty} x\left(-\frac{T}{2}\right) = x_a, \lim_{T \to \infty} x\left(\frac{T}{2}\right) = x_b\} \quad (165)$$

in which $x_a$ and $x_b$ belong to individual attractors.

Now consider the stochastic system with a conserved quantity and Liouville theorem for functionals $\mathcal{K}[\omega]$ and $\mathfrak{F}[\omega]$, repectively:

$$\frac{\partial \omega}{\partial t} = \mathfrak{F}[\omega] \quad (166)$$

$$\nabla \cdot \mathfrak{F} = \int \frac{\delta \mathfrak{F}}{\delta \omega(r)} dr = 0 \quad (167)$$

$$\int \mathfrak{F} \frac{\delta \mathcal{K}}{\delta \omega(r)} dr = 0 \tag{168}$$

Now the Langevin dynamics for the potential $\mathcal{K}[\omega]$ could be defined as [82]:

$$\frac{\partial \omega}{\partial t} = \mathfrak{F}[\omega] - \alpha \int C(r, r') \frac{\delta \mathcal{K}}{\delta \omega(r')} dr' + \sqrt{2\alpha}\, dW_t \tag{169}$$

As enstrophy, energy, and an infinite set of functionals related to vorticity are conserved and $\mathfrak{F}[\omega] = -(u \cdot \nabla)\omega$, it would be simple now to correlate (169) to dynamics of Euler equation. Therefore, formalism developed for Langevin dynamics could be used to analyze non-equilibrium statistical mechanics of Euler equation.

## 5. Numerical methods for stochastic hydrodynamics

In this chapter, some numerical methods exploiting the theory of stochastic differential equations would be presented. Many algorithms have been introduced based on the discretization scheme for stochastic Navier-Stokes equation [83–85], system of forward-backward stochastic differential equations [86–90], or the application of Weiner-chaos expansion [91–94].

### 5.1. System of Forward-Backward Stochastic Differential Equations

It has been demonstrated in Section 3 that it would be possible to derive Navier-Stokes equation using stochastic Lagrangian paths. This implies that it would also be possible to build a system of stochastic differential equations that could be solved coupling with Navier stokes equation.

Consider the following initial value problem:

$$\begin{cases} \frac{\partial}{\partial t} u(t,x) + (u.\nabla)u(t,x) = -\frac{\nabla p}{\rho} + \nu \nabla^2 u(t,x) \\ \nabla . u = 0 \\ u(0,.) = u_0 \end{cases} \quad (170)$$

This could be converted to a control problem using time transformation $t \to T - t$, where T is a fixed positive number. Then setting:

$$\begin{aligned} \tilde{u}(t,x) &= -u(T-t,x) \\ \tilde{p}(t,x) &= p(T-t,x) \end{aligned} \quad (171)$$

For time interval [0,T] it could be concluded that:

$$\begin{cases} \frac{\partial}{\partial t}\tilde{u}(t,x) + (\tilde{u}.\nabla)\tilde{u}(t,x) = -\frac{\nabla p}{\rho} - \nu \nabla^2 u(t,x) \\ \nabla . \tilde{u} = 0 \\ \tilde{u}(T,.) = u_0 \end{cases} \quad (172)$$

Moreover, due to incompressibility condition, pressure obeys Poisson equation and by considering Einstein's notation:

$$\nabla^2 p = \partial_i u^j \partial_j u^i \quad (173)$$

The fundamental solution to (173) can be expressed in terms of the dimension dependent Laplacian Green function $\mathcal{G}_d$ as:

$$p(t,x) = \int \sum_{i=1}^{3} \sum_{j=1}^{3} \partial_i u^j(t,x)\, \partial_j u^i(t,y)\, \mathcal{G}_d(\partial_i u^j \partial_j u^i) \quad , \quad x,y \in \mathbb{R}^2 \quad (174)$$

Using analogy between (172) and (170), (39) could be recast as:

$$\begin{cases} d\tilde{\psi}_s^u(x) = \sqrt{2\nu}\, dW_s + \tilde{u}(s, \tilde{\psi}_s^u)\, ds \\ \tilde{\psi}_0^u(x) = x \end{cases}$$

Now, as $\tilde{u}$ obeys (172):

$$\begin{cases} d\tilde{\psi}_s^u(x) = \sqrt{2\nu}\, dW_s + \tilde{u}(s, \tilde{\psi}_s^u)\, ds \\ d\tilde{u}(s, \tilde{\psi}_s^u) = \nabla \tilde{u}(s, \tilde{\psi}_s^u)\, dW_s + \nabla \mathcal{G}_d(\partial_i \tilde{u}^j \partial_j \tilde{u}^i)\, ds \\ \tilde{u}(T,.) = u_0 \\ \tilde{\psi}_0^u(x) = x \end{cases} \quad (175)$$

As shown, (175) displays a system of forward-backward stochastic differential equations (FBSDE) for which the existence of global [95] and local [96] unique solutions have been proved in two and three-dimensional cases, respectively. It has been suggested that the proposed system could be solved numerically through a four-step scheme [90].

### 5.2. Wiener Chaos Expansion

Prior to discuss the numerical method developed based on WCE, some theoretical prerequisites would be required.

Denote a set of multi-indices with compact support as follows:

$$\mathcal{J} = \left\{ \alpha = (\alpha_1, \alpha_2, \ldots) : \alpha_i \in \mathbb{N}, |\alpha| = \sum_{i=1}^{\infty} \alpha_i < \infty \right\}$$

Let $\xi = (\xi_1, \xi_2, \ldots, \xi_n)$ be a random variable which generates a gaussian measure as:

$$E[f(\xi_1, \xi_2, \ldots, \xi_n)] = \int_{\mathbb{R}^n} f(x_1, \ldots, x_n) \frac{e^{-\frac{1}{2}|x|^2}}{\sqrt{(2\pi)^n}} dx_1 \ldots dx_n = \int_{\mathbb{R}^n} f(x_1, \ldots, x_n)\, d\mu_n(x) \text{ for } f \in L^1(\mu_n)$$

By considering Hermite polynomials as $h_n(x) = (-1)^n e^{\frac{1}{2}x^2} \frac{d^n}{dx^n} e^{-\frac{1}{2}x^2}$, following definitions could be adopted: $H_\alpha(\xi) := \prod h_{\alpha_i}(\xi_I)$, where $\alpha \in \mathcal{J}$

$$\mathfrak{L}^2_{\mu_n} = \oplus_{i=1}^N L^2_{\mu_n}$$

*Theorem 11:* Every $f \in \mathfrak{L}^2_{\mu_n}$ has a unique representation

$$f(\xi) = \sum_{\alpha \in \mathcal{J}} C_\alpha H_\alpha(\xi) \quad \text{where } C_\alpha = E[f(\xi) H_\alpha(\xi)]$$

However, Theorem 11 was generalized for other families of orthonormal polynomials by Xiu and Karniadakis [97]. for example, it could be shown that $W_t = \sum_{i=0}^n \int_0^t m_i(\tau) d\tau \int_0^t m_i(s) dW_s$, in which $\{m_i(s)\}$ is an orthonormal basis for $L^2([0,t])$.

Now, the proposed scheme could be employed to solve the following one-dimensional burgers equation with simplified boundary and initial conditions.

$$\begin{cases} \frac{\partial}{\partial t} u + \frac{1}{2} \frac{\partial}{\partial x} u^2 = \nu \frac{\partial^2}{\partial x^2} u + f(x) \dot{W}(t) \\ u(0, x) = u_0 \\ u(t, 0) = (t, 1) \end{cases} \quad (176)$$

This equation could be expanded in terms of Hermite polynomials:

$$u(x,t) = \sum_{\alpha \in \mathcal{J}} u_\alpha(x,t) H_\alpha(\xi) \quad (177)$$

Also:

$$\begin{cases} u^2(x,t) = \sum_{\alpha \in \mathcal{J}} \sum_{0 \leq \gamma \leq \beta} C(\alpha, \beta, \gamma) u_{\beta-\gamma+\alpha}(x,t) u_{\gamma+\alpha}(x,t) H_\beta(\xi) \\ C(\alpha, \beta, \gamma) = \sqrt{\binom{\beta}{\gamma} \binom{\beta+\alpha}{\alpha} \binom{\alpha+\beta-\gamma}{\alpha}} \end{cases} \quad (178)$$

By substituting (178) and (177) in (176), it could be concluded that:

$$\frac{\partial}{\partial t} u_\alpha(x,t) + \frac{1}{2}\frac{\partial}{\partial x} \sum_{p \in J} \sum_{0 \leq \beta \leq \alpha} C(p,\alpha,\beta)\, u_{\alpha-\beta+p}(x,t) u_{\beta+p}(x,t) =$$

$$\nu \frac{\partial^2}{\partial x^2} u_\alpha(x,t) + f(x) \sum_{i=0}^{\infty} I_{\{\alpha_j = \delta_{ij}\}}\, m_i(t) \tag{179}$$

The obtained system of deterministic PDEs, known as dynamically orthogonal field equations (DOFEs), decouples the stochastic and deterministic part of (176) after expansion. Accordingly, numerical methods for deterministic PDEs could now be employed to find $u_\alpha(x,t)$ coefficients in (177).

However, there are some practical considerations to be addressed. First, (179) is an infinite series that requires two kinds of truncations for a numerical scheme; it should be obviously truncated over $\alpha$ to keep only the Hermite polynomials $H_\alpha, \alpha \leq N$. Furthermore, it should also be truncated over Gaussian random variables of $\xi = (\xi_1, \xi_2, \ldots, \xi_n)$. After such truncations, there exists $\sum_{j=1}^{N} \binom{n+j-1}{j}$ terms in the expansion of (177). However, using the idea of sparse tensor product [98], the number of these terms could considerably decrease. After solving for the coefficients, the moments of $u(x,t)$ could be acquired from following relations:

$$E[u(x,t)] = u_0(x,t)$$

$$E[u^2(x,t)] = \sum_{\alpha \in J} |u_\alpha(x,t)|^2$$

$$E[u^3(x,t)] = \sum_{p \in J} \left( \sum_{\alpha \in J} \sum_{0 \leq \gamma \leq \beta} C(\alpha,\beta,\gamma)\, u_{\beta-\gamma+\alpha}(x,t) u_{\gamma+\alpha}(x,t) \right) u_p(x,t)$$

$$E[u^4(x,t)] = \sum_{p \in J} \left( \sum_{\alpha \in J} \sum_{0 \leq \gamma \leq \beta} C(\alpha,\beta,\gamma)\, u_{\beta-\gamma+\alpha}(x,t) u_{\gamma+\alpha}(x,t) \right)^2$$

Subsequently, it can be seen that $u(x,t)$ is far from a Gaussian process as third and fourth moments are not zero.

The results achieved from this method have been compared to those obtained from Monte Carlo simulations[92,98] and the figures suggest that, with no increase of error, this algorithm would be much more time efficient for first and second moments.

It worth mentioning that, as the time interval becomes larger, the number of Hermite polynomials would also increase, and thereby, a limitation for this method could be long time integrations in the numerical scheme.

Some progresses have been to handle this problem. For instance, Kármán vortex street problem with stochastic frequency has been studied [99] and the method of asynchronous time integration with generalized chaos expansion was used to find limit cycles for such a system. The main idea behind asynchronous time integration is to introduce a rescaled time variable of $\tau(t,\xi)$ such that all realizations $u^\xi(x,t)$ be in phase with some reference solution $u^r(x,t)$ after the transformations $u^\xi(x,t) \rightarrow u(x,\tau(t,\xi))$, which is the constraint to this problem. Applying the chain rule to $\tau(t,\xi)$ results in:

$$\begin{cases} \frac{d}{dt}u(x,\tau(t,\xi)) = \dot{u}((x,\tau(t,\xi)))\frac{d\tau}{dt} \\ u(x,\tau(0,\xi)) = u^r(x,0) = u_0 \end{cases} \qquad (180)$$

To apply the constraint, following criteria could be considered.

$$\Delta(x,t,\xi) = \left(u(x,\tau(t,\xi)) - u^r(x,t)\right).\dot{u}\left((x,\tau(t,\xi))\right) = 0 \qquad (181)$$

If $\Delta < 0$, then $\frac{d\tau}{dt}$ must be greater than zero and vice versa. It could now be deduced that $\frac{d\tau}{dt}$ must obey an equation such as:

$$\begin{cases} \frac{d\dot{\tau}}{dt} = -\alpha\, \Delta(x,t,\xi) + \beta(1-\dot{\tau}) \\ \dot{\tau}(.,0) = 1 \end{cases} \tag{182}$$

Coupling (180) and (182) and expanding $u$ in terms of chaos polynomials leads to a well-defined system of differential equations, which could be solved numerically.

More details including theoretical backgrounds, comparisons between chaos expansion method with and without asynchronous time integration implementations, and computational costs have been reported in the literature [99,100].

## 6. Conclusion

Stochastic theories for Burgers and Navier-Stokes equations were established and existence, uniqueness, and ergodicity were studied for each. Promising results were achieved by the theory of Burgers equation and two-dimensional Navier-Stokes. However, for three-dimensional cases, ergodicity could be proven only in a weak martingale sense and no global result concerning well-Posedness could either be derived. Based on the deterministic variational principle established by Arnold for Euler equation, a stochastic variational principle was also established which gave rise to a system of forward-backward SDEs. Such an approach, known as geometric mechanics, was used to study the group of divergence-free diffeomorphisms.

Burgers turbulence was then studied and dissipative anomaly and anomalous scaling, the features possessed in common by incompressible turbulence, were assessed. Subsequently, Kraichnan model was studied in detail as one of the few solvable models of turbulent transport. Lyapunov exponents and probability density functions were acquired. It was also revealed that Kraichnan model is anomalous and both dissipative anomaly and anomalous scaling were proven to exist in this model. Field theoretic methods were then introduced and employed to solve problems in Burgers and advective turbulence. Such methods were used to handle anomalies and calculating functions of interest, such as probability density and structure functions. Intermittency has also been discussed in three-dimensions and a multifractal model was suggested to evaluate structure functions. It should be mentioned that no feasible strategy has been developed yet to calculate fractal dimensions from the first principles. As an alternative to multifractal model, a hierarchical model was proposed based on the stochastic dynamics of energy cascade. It has been shown that a double cascade paradigm rules in two dimensions and discussed that the two-dimensional turbulence significantly differs from three-dimensional case. Moreover, in two dimensions in which the dissipative anomaly does not exists, presence of a unique random attractor was verified. Inspired by coherent structures and vortex filaments, statistical mechanical methods were developed in case of two-dimensional turbulence. Accordingly, it was shown that such approach could lead to negative temperature prediction. Furthermore, non-equilibrium thermodynamics of turbulence was studied in Fourier space. Another result was non-ergodicity of Euler equation, which motivated the derivation of an instanton theory using Onsager–Machlup formalism to study the transfer rate from one attractor to another.

Ultimately, to approximate the velocity field, two stochastic numerical methods, one a system of forward-backward SDEs based on variational principle, and the other, based on WCE. It has been discussed that WCE provides a framework for calculating moments of velocity and can be used for analytical and numerical purposes. The expansion based on WCE showed non-zero third and fourth moments which verified the non-Gaussian distribution of velocity. In addition, WCE revealed superiority against MCMC simulations in its time consumption. At the end, a strategy for long time calculations of WCE was proposed.

Based on the provided information, it has been shown that turbulence is still far from being solved. No strategy is known for the reconstruction of incompressible turbulence from first principles. However, although there has been progress in two-dimensional case, such advancements do not apply to the three-dimensional turbulence. Stochastic theory of hydrodynamics brings field theoretic machinery into theory of turbulence. The structure provided by random dynamical systems seems natural for describing a complicated theory such as turbulence. It could be concluded that nearly all the major recent developments of turbulence theory strongly depend on probabilistic methods, thereby, a stochastic theory sounds promising. However, such theory is yet far from being complete or self-consistent.

using Reynolds-averaged Navier-Stokes and large-eddy simulation with uncertain inflowconditions, Int. J. Numer. Methods Fluids. 72 (2013) 341–358. https://doi.org/10.1002/fld.3743.

[94] S. Hosder, L. Maddalena, Non-Intrusive Polynomial Chaos for the stochastic CFD study of a supersonic pressure probe, 47th AIAA Aerosp. Sci. Meet. Incl. New Horizons Forum Aerosp. Expo. (2009). https://doi.org/10.2514/6.2009-1129.

[95] F. Delbaen, J. Qiu, S. Tang, Navier-Stokes Equations and Forward-Backward Stochastic Differential Systems, (2013). http://arxiv.org/abs/1303.5329.

[96] X. Chen, A.B. Cruzeiro, Z. Qian, Navier-Stokes equation and forward-backward stochastic differential system in the Besov spaces, (2013). http://arxiv.org/abs/1305.0647.

[97] D. Xiu, G. Em Karniadakis, The Wiener-Askey polynomial chaos for stochastic differential equations, SIAM J. Sci. Comput. 24 (2003) 619–644. https://doi.org/10.1137/S1064827501387826.

[98] P. Frauenfelder, C. Schwab, R.A. Todor, Finite elements for elliptic problems with stochastic coefficients, Comput. Methods Appl. Mech. Eng. 194 (2005) 205–228. https://doi.org/10.1016/j.cma.2004.04.008.

[99] P. Bonnaire, P. Pettersson, C.F. Silva, Intrusive generalized polynomial chaos with asynchronous time integration for the solution of the unsteady Navier–Stokes equations, Comput. Fluids. 223 (2021). https://doi.org/10.1016/j.compfluid.2021.104952.

[100] O.P. Le Maître, L. Mathelin, O.M. Knio, M.Y. Hussaini, Asynchronous time integration for polynomial chaos expansion of uncertain periodic dynamics, Discret. Contin. Dyn.